\documentclass[%
superscriptaddress,
 amsmath,amssymb,
 aps,
 pra, twocolumn
]{revtex4-2}

\usepackage{graphicx}
\usepackage{dcolumn}
\usepackage{bm}
\usepackage{xcolor}

\newcommand{\prob}{{\rm pr}}

\begin{document}

\title{Noise resilience of deterministic analog combinatorial optimization solvers}

\author{Clemens Gneiting}
 \email{clemens.gneiting@riken.jp}
\affiliation{Center for Quantum Computing (RQC), RIKEN, Wako, Saitama 351-0198, Japan}
\affiliation{Theoretical Quantum Physics Laboratory, Cluster for Pioneering Research, RIKEN, Wako, Saitama 351-0198, Japan}

\author{Farad Khoyratee}
\affiliation{Theoretical Quantum Physics Laboratory, Cluster for Pioneering Research, RIKEN, Wako, Saitama 351-0198, Japan}
\affiliation{Department of Physics, University of Michigan, Ann Arbor, Michigan 48109-1040, USA}
 
\author{Enrico Rinaldi}
\affiliation{Quantinuum K.K., Otemachi Financial City Grand Cube 3F, 1-9-2 Otemachi, Chiyoda-ku, Tokyo, Japan}
\affiliation{Center for Quantum Computing (RQC), RIKEN,	Wako, Saitama 351-0198, Japan}
\affiliation{Theoretical Quantum Physics Laboratory, Cluster for Pioneering Research, RIKEN, Wako, Saitama 351-0198, Japan}
\affiliation{Center for Interdisciplinary Theoretical and Mathematical Sciences (iTHEMS), RIKEN, Wako, Saitama 351-0198, Japan}

\author{Khyati Jain}
\affiliation{Department of Physics, BITS Pilani, Goa Campus, Goa, India}
\affiliation{Theoretical Quantum Physics Laboratory, Cluster for Pioneering Research, RIKEN, Wako, Saitama 351-0198, Japan}

\author{Rishab Khincha}
\affiliation{Department of Physics, BITS Pilani, Goa Campus, Goa, India}
\affiliation{Theoretical Quantum Physics Laboratory, Cluster for Pioneering Research, RIKEN, Wako, Saitama 351-0198, Japan}
\affiliation{The University of Texas at Austin, USA}

\author{Franco Nori}
\affiliation{Center for Quantum Computing (RQC), RIKEN, Wako, Saitama 351-0198, Japan}
\affiliation{Theoretical Quantum Physics Laboratory, Cluster for Pioneering Research, RIKEN, Wako, Saitama 351-0198, Japan}
\affiliation{Department of Physics, University of Michigan, Ann Arbor, Michigan 48109-1040, USA}

\date{\today}

\begin{abstract}
Several continuous dynamical systems have recently been proposed as special-purpose analog computers designed to solve combinatorial optimization problems such as $k$-SAT or the Ising problem. While combinatorial optimization problems are known to be NP-hard, and thus scale, in the worst case, exponentially with the problem size, these analog solvers promise substantial speed-up and scaling advantages in finding the solution. The underlying algorithms, which can be cast in the form of differential equations, generically involve highly chaotic dynamics and thus assume that the system variables can be processed with, in principle, arbitrary precision. However, both actual physical systems as well as finite digital machines, which are used to virtually emulate the dynamics, can process the evolution only with finite precision, be it because of intrinsic noise or because of limited precision in number representation. We investigate the impact of such noise on the solution-finding capability. To this end, we focus on two representative analog solvers, designed to address the Ising problem and the $k$-SAT problem, respectively. Our numerical analysis reveals that the ability of these algorithms to find solutions exhibits a threshold behavior under the addition of noise, where the solution-finding capability remains mostly uncompromised below a noise threshold, while it rapidly deteriorates above the threshold. As we show, these noise tolerance thresholds decrease with the problem size, following an approximate algebraic scaling. This allows us to infer principal limits on the problem sizes that can be efficiently tackled with these solvers under given noise levels.
\end{abstract}

\maketitle

\section{Introduction}

Combinatorial optimization lies at the heart of games like chess and sudoku, academic puzzles like the shortest path for a traveling salesman, and fundamental scientific challenges such as the question of logical satisfiability ($k$-SAT), or the determination of the ground-state energy of paradigmatic condensed-matter models like the Ising chain \cite{Papadimitriou1998combinatorial}. In addition, many practical optimization tasks in modern society, such as real-time traffic control or network analysis and design, take the form of combinatorial optimization problems. Often these practical usecases can be naturally mapped onto a representative subset of combinatorial optimization problems that prominently includes the Ising and the $k$-SAT problem classes \cite{Lucas2014ising}.

In view of their growing relevance, identifying efficient solution strategies for combinatorial optimization problems has become of paramount importance to advance the scope and scale of their applicability. This progress is mainly hindered by the exponential growth of the search space with problem size, which rapidly renders a naive brute-force exploration of all possibe combinations prohibitive, and which identifies combinatorial optimization problems as core members of the NP-hard computational complexity class.

Physics-inspired stochastic solution strategies have long been successfully deployed to circumvent the need of a full exploration of the search space \cite{Schoening1999probabilistic}. For instance, {\it simulated annealing} algorithms interpret combinatorial optimization problems in terms of physical systems, with the solution heuristics that slow cooling drives them into their ground states \cite{Bertsimas1993simulated} (for a deterministic take on simulated annealing, see~\cite{Chen1995chaotic}).

More recently, (classical) analog computational approaches have been in the focus of intense research, either inspired by concrete physical systems such as the Coherent Ising Machine, or with the idea to take advantage of the increased flexibility that comes with originally discrete variables that have been ``softened'' to continuous variables \cite{Ercsey2011optimization, Ercsey2012chaos, Wang2013coherent, Yamamoto2017coherent, Berloff2017realizing, King2018emulating, Leleu2019destabilization, Goto2019combinatorial, Vadlamani2020physics, Goto2021performance, Mohseni2022ising, Reifenstein2023coherent, Syed2023physics}. Many of these analog systems are deterministic, which suggests---and in some cases it can be proved---that a solution is found with certainty. Indeed, the expectation that analog algorithms may unlock new computational capabilities is supported by preceeding results that show, for instance, that a computing machine with the capability to multiply real numbers with arbitrary precision would empower this machine to solve satisfiability problems in polynomial time \cite{Schoenhage1979power}.

Deterministic analog approaches to combinatorial optimization replace a brute-force search, or a stochastic exploration of the search space, by chaotic dynamics. The underlying physical intuition is that chaotic systems are ergodic and hence explore their phase space exhaustively, eventually finding a solution. However, it is well known that chaotic dynamics are highly sensitive to perturbation, be it to the initial state or at any later stage of the evolution. This sensitivity is also reflected by the scale volatility of chaotic systems, suggesting that microscopic perturbations can rapidly inflate to macroscopic deviations (``butterfly effect'').

The sensitivity of deterministic analog combinatorial optimization solvers to perturbations raises the natural question if or to what extend they can tolerate noise without losing their solution-finding capability. The relevance of this question is manifest when such an algorithm is run on actual analog physical systems, where finite noise levels are unavoidable. But, maybe less obviously, this question is also relevant in digital emulations of these algorithms, where efficiency often comes at the price of reduced precision in the number representation; as it is, for instance, the case with field-programmable gate arrays (FPGAs).

We address the question of the noise sensitivity of deterministic analog combinatorial optimization solvers for two prominent members of this class. The first algorithm, inspired by the Coherent Ising machine, is designed to solve the Ising problem and has been shown in benchmark tests to exhibit promising solution-finding and scaling properties \cite{Leleu2019destabilization, Leleu2021scaling}. The second algorithm addresses $k$-SAT problems and has been proven to find a solution with certainty \cite{Ercsey2011optimization, Ercsey2012chaos}. Moreover, there is strong numerical evidence that, given arbitrarily precise number representation, it does so in polynomial (system) time. While both these algorithms give rise to (transiently) chaotic dynamics, their solution-finding strategies are different, so that we take them as a generic, representative sample of the class of deterministic analog combinatorial optimization solvers.

We find that the solution-finding capability of both algorithms exhibits a threshold behavior under the exposure of noise. That is, the solution-finding capability is only marginally (however with a qualification for the $k$-SAT solver, as explained below) affected below the noise threshold, while it rapidly decays above the threshold. As we show, these noise-tolerance thresholds decrease with the problem size and approximately follow an algebraic scaling. We use this to infer principal limits on the problem sizes that can be efficiently addressed with these solvers under given noise levels. For investigations of the noise impact on deterministic analog solvers with different foci see, for instance, \cite{Albash2019analog, Sumi2014robust}.

This article is structured as follows: In Section~\ref{Sec:Analog_Ising_solver}, we present the analog Ising solver, the class of Ising problems that we consider, our noise model, and our numerical analysis that demonstates the emergence of noise thresholds and their scaling with $N$. We show this for dense and sparse interaction matrices, and for white and colored noise. In Section~\ref{Sec:Analog_k-SAT_solver}, we analogously discuss the analog $k$-SAT solver. An interesting feature we will encounter here is the emergence of a ``soft noise threshold'', that is, an ($N$-independent) threshold that filters an ($N$-dependent) fraction of problem instances from being solved. This soft noise threshold is featured in addition to the regular ``hard noise threshold'' that filters all problem instances. In Section~\ref{Sec:Discussion}, we then use the obtained noise threshold scalings to predict noise-dependent limits of applicability, followed by the concluding Section~\ref{Sec:Conclusions}.

\section{Analog Ising solver} \label{Sec:Analog_Ising_solver}

In this section, we investigate the noise impact on the {\it chaotic amplitude control} (CAC) algorithm \cite{Leleu2019destabilization, Leleu2021scaling}. The CAC algorithm has been proposed as an efficient analog solver for the Ising problem. The Ising problem, while physically motivated, is a generic NP-hard combinatorial optimization problem that can be related to many practical optimization tasks. Whereas such optimization tasks play an increasingly important role in a digital society, brute force solution-finding strategies become rapidly prohibitive, highlighting the relevance of efficient alternative approaches.

\subsection{The Ising problem}

The Ising model was initially conceived as a minimalistic model to describe ferromagnets. It consists of $N$ binary degrees of freedom $\sigma_i \in \{-1,1\}$, $i \in \{1,\dots,N\}$, ``spins'', which are mutually coupled through an interaction matrix $W_{ij}=W_{ji}$ \cite{Nishimori2001statistical}. The energy of an arbitrary spin configuration is then given by the Hamiltonian
\begin{align} \label{Eq:Ising_Hamiltonian}
	H = -\frac{1}{2} \sum_{i,j =1}^N W_{ij} \sigma_i \sigma_j .
\end{align}
The Ising problem consists of finding the ground state, i.e., for a given interaction matrix $W_{ij}$, finding the spin configuation which minimizes the energy (\ref{Eq:Ising_Hamiltonian}). Note that, if $W_{ij}>0$ for fixed $i$ and $j$, then coaligned spins $\sigma_i$ and $\sigma_j$ minimize the energy of the bond between the spins $i$ and $j$, while, if $W_{ij}<0$, antialigned spins minimize the energy of the bond. The fact that one can in general not minimize the energy of all bonds simultaneously is referred to as {\it frustration} and gives rise to the computational hardness of the Ising problem.

Restrictions on the choice of the interaction matrix elements $W_{ij}$ are usually motivated by the underlying physical or optimization problem. A common assumption is, for instance, that the bonds $W_{ij}=W_{ji}$ are randomly and independently drawn from a Gaussian distribution (Sherrington-Kirkpatrick model), which allows analytical mean-field treatments. For our purpose, the most relevant model property is the problem hardness that comes with frustration, and therefore we focus on randomly and independently drawn interaction matrix elements
\begin{align} \label{Eq:dense_discrete_Ising_interaction}
	W_{ij} \in \{-1,1\} .
\end{align}
Note that our interaction matrices are dense, i.e., bonds are nonvanishing between every pair of spins. At the end of this section, we also briefly discuss the case of sparse interaction matrices.

\subsection{Optically-inspired solution-finding approach}

An exhaustive brute-force exploration to find the ground state by calculating (\ref{Eq:Ising_Hamiltonian}) for all $2^N$ possible spin configuations becomes rapidly prohibitively computationally expensive, and thus does not present a viable solution-finding strategy for practical problems. Therefore, one has traditionally resorted to probabilistic methods such as simulated annealing \cite{Bertsimas1993simulated}. The underlying intuition here is that a slow reduction of the (simulated) temperature drives the spin configuration towards the ground state, while ``bath''-induced stochastic perturbations help to avoid that the state gets trapped in local minima of the potential landscape.

An alternative approach has recently been proposed in the context of networks of coupled degenerate optical parametric oscillators (DOPOs) \cite{Wang2013coherent, Marandi2014network, Mcmahon2016fully, Inagaki2016coherent, Yamamoto2017coherent, Yamamoto2020coherent, Honjo2021spin, Zhou2023coherent, Zhou2024frustration}. The core idea underlying this approach is that DOPOs undergo a phase bifurcation once the pump strength exceeds a threshold value, suggesting the identification with an effective spin. Specifically, the field amplitudes are regarded as ``soft spins'', while the sign values of the asymptotic steady state amplitudes determine the final spin assignments. The heuristic expectation then is that, if the DOPOs are coupled according to the Ising interaction $W_{ij}$, and if the pump is adiabatically ramped up from (well) below to (well) above the collective threshold, then the Ising ground state is excited. Hence, in contrast to simulated annealing, where the ground state is {\it probabilistically} approached ``from above'', the ground state here is {\it deterministically} approached ``from below''.

The classical (mean-field) approximation of the dynamics of the Ising-coupled DOPOs is given by
\begin{align} \label{Eq:classical_CIM}
	\frac{dx_i}{dt} &= [-1 + p(t)] x_i - x_i^3 + \epsilon e_i \sum_{j} W_{ij} x_j ,
\end{align}
where the $x_i \in \mathbb{R}$ denote the field amplitudes (soft spins), $p(t)$ describes the time-dependent pump, $\epsilon$ stands for a global coupling strength, and the ``error variables'' (the name will become clear below) $e_i$ can be regarded as an additional modulation of the individual coupling of each spin to all other spins. According to the intuition presented above, a natural choice is $e_i = e = 1$, that is, the coupling among the soft spins is entirely modulated by the Ising interaction while the global parameter $\epsilon$ controls the strength of how the interaction modifies the local DOPO dynamics.

In the simplest implementation, the pump $p(t)$ is ramped linearly from below to above the threshold. Note that we describe time $t$ in units of $\gamma^{-1}$, where $\gamma$ denotes the rate of optical losses occurring in the DOPOs. In this convention, isolated DOPOs ($\epsilon=0$) cross the threshold if $p(t)>1$, that is, if the pump (over)compensates the optical losses. This is easily verified with the DOPO potential $V_t(x) = -\frac{1}{2}(-1+p(t)) x^2 + \frac{1}{4} x^4$, which becomes bistable for $p>1$, with two stable minima at $\pm \sqrt{p-1}$.

The approach outlined above has been successfully applied to the Ising problem with Coherent Ising Machines (CIMs) \cite{Yamamoto2017coherent, Mohseni2022ising, Reifenstein2023coherent}, where a train of optical pulses representing the soft spins travels through a loop cavity that contains a nonlinear crystal as an amplification element. In present-day scalable CIMs, the couplings between the pulses are realized by cyclic homodyne measurements followed by FPGA-processed feedback injections, a coupling scheme that is captured well by the classical description (\ref{Eq:classical_CIM}). Using this approach, specific classes of Ising problems with thousands of spins have been successfully tackled \cite{Inagaki2016coherent, Honjo2021spin}.

\subsection{Chaotic amplitude control algorithm}

Despite the remarkable performance of CIMs in solving Ising problems, it is well known that the above outlined adiabatic approach does in general not deliver the correct Ising ground state, irrespective of how well the adiabatic condition is met by the pump ramp \cite{Kako2020coherent}. Indeed, it can be shown that the first state that is excited when crossing the collective threshold is the eigenstate of the interaction matrix $W_{ij}$ corresponding to the smallest eigenvalue \cite{Yamamura2024geometric}. While the sign configuration of this eigenstate is often a good approximation to the exact ground state, it does in general not correspond to the exact eigenstate.

The in-general failure of the adiabatic algorithm can be traced back to the generic amplitude heterogeneity assumed by the soft spins in the steady states. Indeed it can be shown that, if all amplitudes maintain the same magnitude (not sign), the true Ising ground state is reached \cite{Leleu2017combinatorial}. Following this logic, a modification of the above algorithm has been proposed, where the error variables $e_i$ in (\ref{Eq:classical_CIM}) are now dynamic and designed to keep all amplitudes $|x_i|$ close to a common time-dependent amplitude $a(t)$ \cite{Leleu2019destabilization}. Specifically, the evolution of the error variable $e_i$ is described by
\begin{align} \label{Eq:CAC_error_variables_dynamics}
	\frac{de_i}{dt} &= -\beta(t) [x_i^2 - a(t)] e_i ,
\end{align}
where the global $\beta(t)$ provides additional freedom to control the strength of this ``error correction'' mechanism. One easily sees that the dynamics (\ref{Eq:CAC_error_variables_dynamics}) decreases (increases) the magnitude of the error variable whenever the corresponding (squared) soft spin variable $x_i^2$ overshoots (undershoots) the target amplitude $a(t)$, thus steering the impact of the interactions on the dynamics of the  spin variable, cf.~Eq.~(\ref{Eq:classical_CIM}).

The family of analog spin solvers that are characterized by the dynamics (\ref{Eq:classical_CIM}) and (\ref{Eq:CAC_error_variables_dynamics}) has been dubbed {\it chaotic amplitude control} algorithms, as the resulting spin dynamics is generically chaotic. We focus here on a specific member of this family, which has shown promising benchmarking behavior \cite{Leleu2021scaling}. It is defined through
\begin{subequations} \label{Eq:CAC_specifics}
	\begin{align}
		p(t) &= -\pi + \rho_p \text{tanh}\left(\delta_p\Delta\mathcal{H}(t)\right) \\
		a(t) &= -\alpha + \rho_a \text{tanh}\left(\delta_a\Delta\mathcal{H}(t)\right) \\
		\Delta\mathcal{H}(t) &= \mathcal{H}_{\text{opt}} - \mathcal{H}(t) \\
		\frac{d\beta(t)}{dt} &= \begin{cases} \gamma,& \text{if } t - t_r \leq\tau/\Delta t \\ 0, & \text{otherwise} \end{cases} .
	\end{align}
\end{subequations}
We choose the parameters as in \cite{Leleu2021scaling}, where they have been optimized for performance with respect to the problem class (\ref{Eq:dense_discrete_Ising_interaction}). Specifically, the pump parameters are given by $\pi=-1.82$, $\rho_p=0.00$, and $\delta_p=0.00$, keeping the pump effectively constant (above threshold). The parameters for the target amplitude are set as $\alpha=1.00$, $\rho_a=3.00$, and $\delta_a=4.00$, with the initial condition $a(0)=1.00$. $\mathcal{H}(t)$ denotes the Ising energy of the spin configuration at time $t$, while $\mathcal{H}_{\text{opt}}$ stands for the energy of the current best solution candidate. Similar to above, the sign patterns of the soft spins $x_i(t)$ are taken as the current spin configuration. When the best solution candidate is updated at time $t_r$, then $\beta(t)$ grows linearly with a rate $\gamma=5.5 \times 10^{-4}$ for at most a duration $\tau=900$, starting at $\beta(0)=0.00$. We use Euler's method to numerically solve the differential equations with a constant time step $\Delta t=2^{-6}$ throughout all of our simulations. The initial conditions for the error variables are $e_i(0)=1.00$, and the global interaction strength is chosen as $\epsilon=0.27$. The initial conditions for the soft spins are drawn randomly from a homogeneous distribution in the interval $[-1,1]$. At a problem size-dependent cutoff time $T_{\rm cut}=N^2$, that is, after a total of $\frac{N^{2}}{\Delta t}$ evolution steps, the evolution is terminated. Note that it is not required that the system has reached a steady state at $T_{\rm cut}$.

Let us emphasize that, while we choose the specific algorithm defined by (\ref{Eq:CAC_specifics}) for its good benchmarking performance, we do not expect that these details of the algorithm have a strong impact on its noise response. For our purpose, we consider it as a generic example displaying a noise response that should, at least qualitatively, be representative for the entire family of CAC algorithms.

\subsection{Noise model}

The characteristics of the noise depend on the specific physical or digital implementation of the CAC algorithm. For instance, in the case of DOPO networks intrinsic quantum noise ultimately limits the signal-to-noise ratio of the pulses. In digital computing machines, finite number grids effectively introduce noise by randomizing subprecision digits. In field-programmable gate arrays (FPGAs), for example, it is common to represent numbers with fixed-digit precision.

Our goal here is to investigate the generic noise impact on the solution-finding capability of the CAC algorithm, independent of its specific implementation. To this end, we add controllable, artificial noise to the evolution equations (\ref{Eq:classical_CIM}) and (\ref{Eq:CAC_error_variables_dynamics}), such that the artificial noise dominates over the intrinsic noise of the digital simulation. As a generic, implementation-agnostic noise model, we assume Gaussian white noise with variable intensity $D$,
\begin{subequations} \label{Eq:CAC_with_noise}
	\begin{align}
		\frac{dx_i}{dt} =& [-1 + p(t)] x_i - x_i^3 + \epsilon e_i \sum_{j} W_{ij} x_j +\sqrt{2D} \xi_i^{(\sigma)}(t) \\
		\frac{de_i}{dt} =& -\beta(t) [x_i^2 - a(t)] e_i + \sqrt{2D}\xi_i^{(e)}(t) ,
	\end{align}
\end{subequations}
where $\xi_i^{(\sigma)}(t) = \mathcal{N}(0, \Delta t^{-1})$ and $\xi_i^{(e)}(t) = \mathcal{N}(0, \Delta t^{-1})$ denote normal random variables with vanishing mean and variance $1/\Delta t$. Recall that we implement the CAC algorithm with a fixed step size $\Delta t = 2^{-6}$, and hence white noise corresponds to statistically independent noise kicks at each time step, $\langle \xi_i^{(\sigma)}(t) \xi_i^{(\sigma)}(t+\Delta t) \rangle = \langle \xi_i^{(e)}(t) \xi_i^{(e)}(t+\Delta t) \rangle = 0$.

By varying the noise intensity, we can analyze the noise impact on the solution-finding capability depending on the problem size (and hardness). For this purpose, we can expect that other noise models, e.g. colored and/or non-Gaussian noise, while different in the details of their effect on the evolution, give rise to similar (statistical) noise responses. Below we verify this for temporally-correlated noise.

\subsection{Data generation}

\begin{figure*}[htb]
	\centering
	\includegraphics[width=0.99\textwidth]{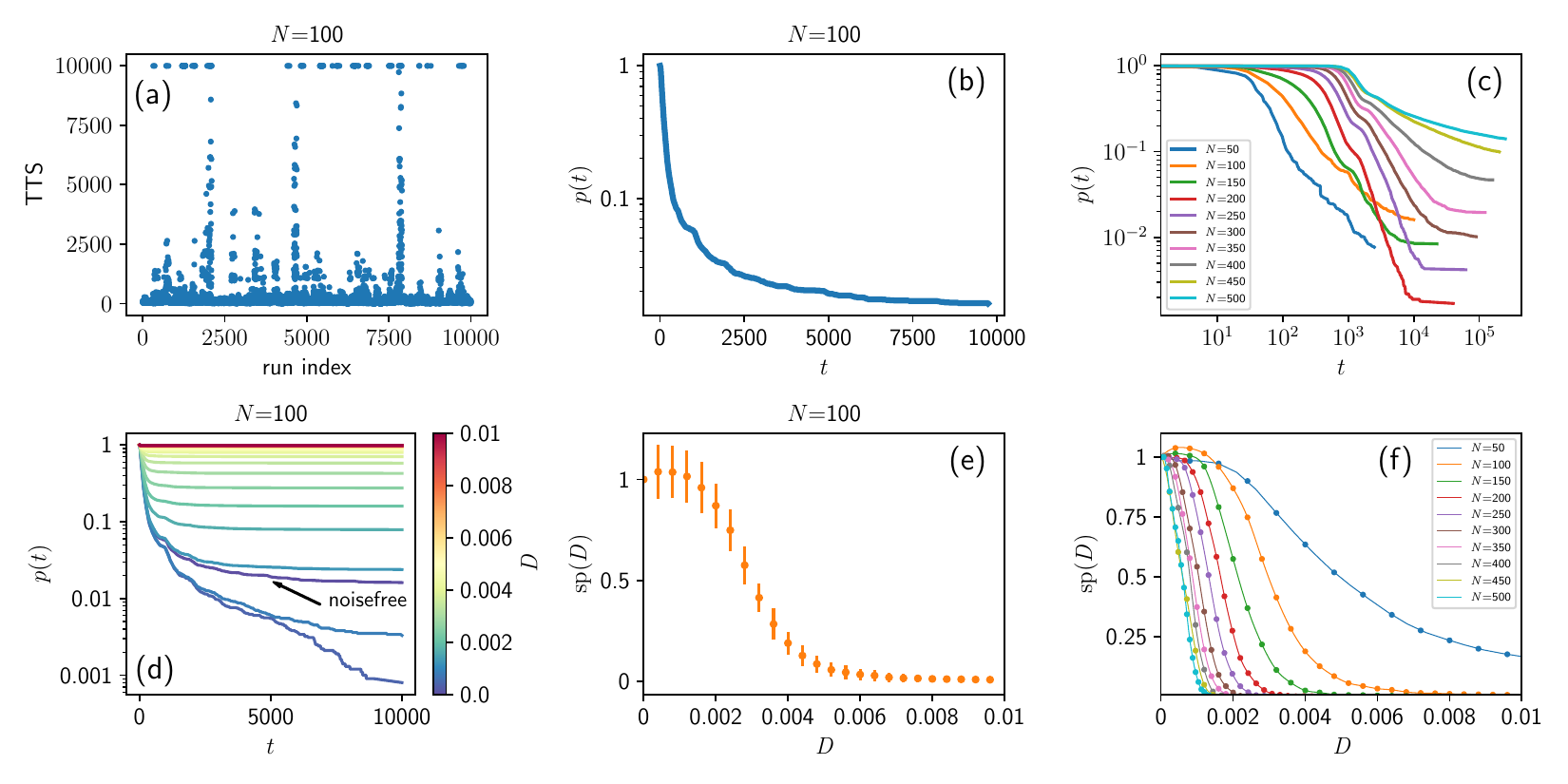}
	\caption{\label{Fig:CAC_data_processing} Data generation and processing for the chaotic amplitude (CAC) algorithm, and the emergence of noise thresholds. (a) Time-to-solution (TTS) data, obtained with the CAC algorithm (\ref{Eq:classical_CIM}, \ref{Eq:CAC_error_variables_dynamics}, \ref{Eq:CAC_specifics}) in the absence of noise, for $100$ random Ising problem instances (\ref{Eq:dense_discrete_Ising_interaction}) of size $N=100$. Each problem instance is run with $100$ random initial conditions. If the algorithm does not find a solution candidate that reaches or undershoots $\mathcal{H}_{\text{opt},0}$ before the cutoff time $T_{\rm cut}=N^2=10^4$, $T_{\rm cut}$ is assigned as TTS. (b) Fraction of unsolved problems $p(t)$ for the same data. By definition, $p(t)$ monotonically decreases from $p(0)=1$. At the cutoff time $T_{\rm cut}$ about $1.6\%$ of the runs have not reached a solution. (c) Fraction of unsolved problems $p(t)$ for all problem sizes from $N=50$ to $N=500$. With increasing $N$ the $p(t)$ flattens, reflecting increasing problem hardness on average. (d) Fraction of unsolved problems $p(t)$ for $N=100$ in the presence of noise, cf.~Eqs.~(\ref{Eq:CAC_with_noise}). While small noise intensities $D$ accelerate the decay of $p(t)$ and hence enhance the solution-finding capability, $p(t)$ assumes, for sufficiently large $D$, constant asymptotic values, implying that the corresponding fraction of problems becomes unsolvable. (e) Success probability ${\rm sp}(D)$ as a function of the noise intensity $D$. The transition from a small-noise regime, where noise does not have a detrimental effect, to a regime, where noise prohibits any solution finding, establishes a noise threshold. (f) Success probabilities ${\rm sp}(D)$ for all problem sizes from $N=50$ to $N=500$. The noise threshold moves to smaller noise values with increasing problem size. The data points are connected by a spline interpolation to guide the eye.}
\end{figure*}

In order to access the statistical noise effect on the CAC algorithm, we generate, for each problem size $N$, 100 random problem instances of the type (\ref{Eq:dense_discrete_Ising_interaction}). Each problem instance is evaluated for an $N$-dependent range of noise intensities $D$. Each problem instance and noise intensity is run with 100 different randomly drawn sets of spin initial conditions $x_i(0) \in [-1,1], i \in \{1, \dots, N\}$. In total we obtain statistics from $N_{\rm run}=10^4$ runs per problem size and noise intensity. The problem sizes span one order of magnitude, from $N=50$ to $N=500$, in steps of 50.

Since we do not know the exact solutions to the Ising problem instances for the examined problem sizes, we consider a run in the presence of noise successful (in the sense that a valid solution is found) if the energy $\mathcal{H}_{\text{opt}}$ of the optimal solution candidate, within the cutoff time $T_{\rm cut}$, reaches or drops below the optimal energy $\mathcal{H}_{\text{opt},0}$ found in the absence of noise. $\mathcal{H}_{\text{opt},0}$ is chosen to be the smallest energy found among 100 separately randomly drawn sets of spin initial conditions.

In each run we record the time-to-solution (TTS), defined as the first time when $\mathcal{H}_{\text{opt}}$ reaches or undershoots the $\mathcal{H}_{\text{opt},0}$ of the corresponding problem instance. If $\mathcal{H}_{\text{opt}}$ remains above $\mathcal{H}_{\text{opt},0}$ until the cutoff time $T_{\rm cut}$ is reached, then we consider the run as failed and assign $T_{\rm cut}$ as TTS. Note that failure can, depending on the initial conditions, also occur in the absence of noise. In~Fig.~\ref{Fig:CAC_data_processing}(a) we show the noise-free TTS distribution for $N=100$, where bins of 100 refer to different initial conditions for the same problem instance. One nicely sees how some hard problem instances tend to require larger TTS. Moreover, we find that several initial conditions reach $T_{\rm cut}$ without finding an eligible solution.

For our purposes, the most informative presentation of the TTS data is in terms of the fraction of unsolved problems at time $t$,
\begin{align} \label{Eq:fraction_of_unsolved_problems}
	p(t) = 1- \frac{N_{\rm sol}(t)}{N_{\rm run}} ,
\end{align}
where $N_{\rm sol}(t)$ describes the number of runs that have reached or undershot the target energy before time $t$. By definition, this is a monotonically decreasing function with $p(0)=1$. Figure~\ref{Fig:CAC_data_processing}(b) shows $p(t)$ for the TTS data in Fig.~\ref{Fig:CAC_data_processing}(a). We find that, at the cutoff time $T_{\rm cut}=N^2=10^4$, $p(t)$ has dropped to $p(T_{\rm cut}) \approx 0.016$, reflecting that about $1.6\%$ of the runs have not found a solution before reaching $T_{\rm cut}$.

Similarly, we obtain the fraction of unsolved problems $p(t)$ for all other problem sizes, as shown in Fig.~\ref{Fig:CAC_data_processing}(c). We generally observe that the decay of $p(t)$ flattens out with increasing $N$, reflecting the increasing problem hardness. Nevertheless, even at the largest problem size, $N=500$, the majority of runs finds a viable solution, evidencing the solution-finding power of the CAC algorithm if noise is negligible. We note that the three smallest problem sizes ($N=50, 100, 150$), in the long-time limit, break the otherwise monotonic scaling behavior. This may be because these still transition into the asymptotic-in-$N$ scaling behavior.

Adding noise to the evolution equations, as described by Eqs.~(\ref{Eq:CAC_with_noise}), can substantially modify the decay of the fraction of unsolved problems $p(t)$. This is shown in Fig.~\ref{Fig:CAC_data_processing}(d) for a span of noise intensities $D$ for $N=100$. While small noise intensities have a beneficial effect and tend to decrease the TTS, sufficiently large noise intensities give rise to a persistent fraction of unsolvable problems, as reflected by constant asymptotic values of $p_D(t)$. Importantly, we observe that the fraction of unsolvable problems $p_D(T_{\rm cut})$ increases with increasing $D$.

\subsection{Noise threshold scaling}

The noise-induced emergence of a fraction of unsolvable problems motivates us to define the success probability in the presence of noise,
\begin{align} \label{Eq:CAC_success_probability}
	{\rm sp}(D) =\frac{1 - p_D(T_{\rm cut})}{1 - p_{D=0}(T_{\rm cut})}.
\end{align}
Note that, in order to isolate noise effects, we renormalize the success probability to be $1$ in the absence of noise ($D=0$). In Figure~\ref{Fig:CAC_data_processing}(e) we show ${\rm sp}(D)$ for our data for $N=100$. We first observe that, for small noise intensities, ${\rm sp}(D)$ grows larger than $1$, again indicating that these noise intensities, for this problem size, are indeed beneficial for the solution finding. However, for noise intensities beyond $D \approx 0.002$, ${\rm sp}(D)$ starts to drop drastically, until it reaches a vanishing ${\rm sp} \approx 0$ at around $D \approx 0.006$. We interpret this transition from a small-noise regime, where the noise has negligible (in the present case, no detrimental) impact, to a regime where the noise prohibits any solution finding, as the manifestation of a noise threshold. The error bars in Fig.~\ref{Fig:CAC_data_processing}(e) describe the variance of the distribution of ${\rm sp}(D)$ over different problem instances. The decreasing variance indicates that different problem instances behave increasingly uniformly with increasing $D$.

Similar noise thresholds emerge for larger $N$, as well, as shown in Fig.~\ref{Fig:CAC_data_processing}(f). We observe that the noise thresholds consistently decrease with increasing $N$, together with the widths of the transitions. Note that the benefit of small noise for the solution finding, indicated by ${\rm sp}(D)>1$, disappears for larger $N$. The smallest problem size $N=50$ presents itself as an outlier, in the sense that it does not yet display a clear threshold behavior. Again, we explain this through the yet incomplete transition into the typical large-$N$ behavior. This outlier role of $N=50$ will reoccur in the scaling analysis of the noise threshold.

\begin{figure*}[htb]
	\centering
	\includegraphics[width=0.99\textwidth]{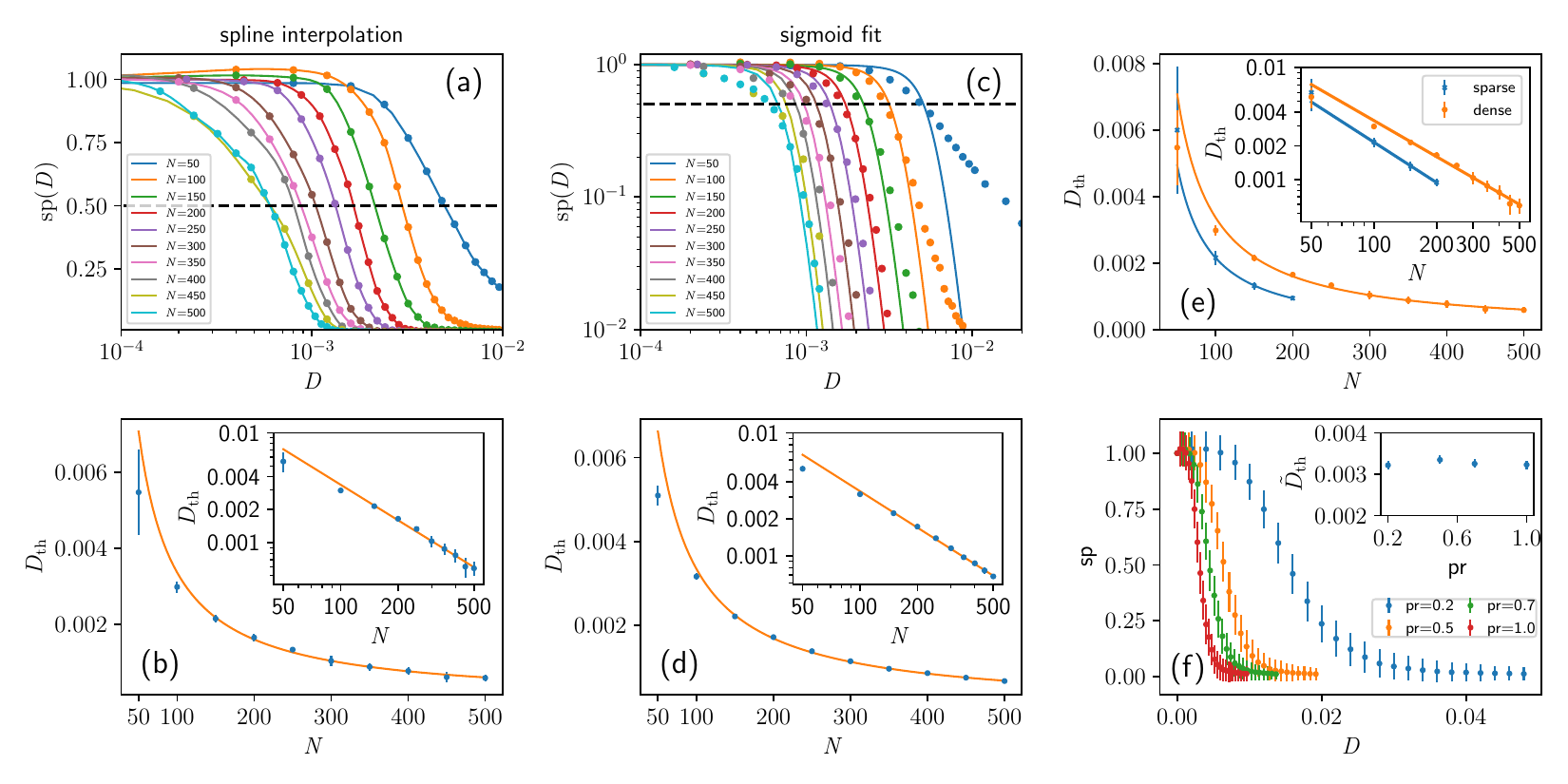}
	\caption{\label{Fig:CAC_noise_threshold_scaling} Noise threshold scaling with problem size $N$ for the chaotic amplitude control (CAC) algorithm. We use both (a) spline interpolations and (c) sigmoid fits to extract the noise threshold values $D_{\rm th}$, defined by ${\rm sp}(D_{\rm th})=0.5$ (blue solid horizontal lines). (b, d) The extracted threshold values [blue dots in (b, d)] follow a polynomial asymptotic scaling with $N$, where (b) spline interpolation and (d) sigmoid fit deliver comparable scaling predictions. The insets show the same data points in double-logarithmic representation. The orange solid lines represent best polynomial fits of the form $a N^b$, where $a \approx 0.46$ and $b \approx -1.07$ (spline interpolation), and $a \approx 0.38$ and $b \approx -1.02$ (sigmoid fit), respectively. Small $N$ values still transition into the asymptotic scaling behavior. (e) Noise threshold scaling for sparse interaction matrices. Shown are the noise thresholds for the sparse interaction matrices (\ref{Eq:sparse_discrete_Ising_interaction}) (blue stars) together with the noise thresholds of the dense interaction model (\ref{Eq:dense_discrete_Ising_interaction}) [orange dots, as in (d)]. The solid lines represent the respective best fits and the inset shows the same in double-logarithmic representation. We find that the polynomial scaling behavior is maintained under sparsity, with a shift to smaller noise values and a marginally larger (more negative) fitting exponent $b$, cf.~(\ref{Eq:CAC_sparse_noise_threshold_scaling_law}). (f) Noise threshold behavior with colored noise for $N=100$. If the noise kicks occur with a probability $\prob<1$ at each time step, the emerging noise thresholds are shifted to larger noise strengths $D$ per noise kick. On the other hand, the rescaled noise thresholds $\tilde{D}_{\rm th}=D_{\rm th} \times \prob$ are invariant under the choice of $\prob$ (inset), indicating the robustness of the emerging noise thresholds under variations of the noise model.}
\end{figure*}

To quantify the noise threshold behavior, we define the noise threshold $D_{\rm th}$ as the midpoint, where the success probability has dropped to $50\%$, that is, ${\rm sp}(D_{\rm th})=0.5$. In Figure~\ref{Fig:CAC_noise_threshold_scaling}(a), we use spline interpolations to connect our ${\rm sp}(D)$ data, which allows us to conveniently estimate $D_{\rm th}$ for all $N$. Indeed, we find that we can also quite accurately extract the $D_{\rm th}$ values by fitting the transitions with sigmoids, ${\rm sp}(D) = 1/[1-\exp(-(D-D_{\rm th})/\Delta D)]$ (with transition width $\Delta D$), as shown in Fig.~\ref{Fig:CAC_noise_threshold_scaling}(c). In the following we use both extraction methods (spline interpolation and sigmoid fit) to infer the noise threshold scaling with $N$.

The noise thresholds rapidly converge towards a regular, polynomial scaling behavior with $N$. This is shown in Fig.~\ref{Fig:CAC_noise_threshold_scaling}(b,d), where the noise thresholds from our data (blue dots) are plotted together with a polynomial best fit (orange solid line) of the form
\begin{subequations} \label{Eq:CAC_noise_threshold_scaling_law}
\begin{align}
	D_{\rm th}(N) = a N^b .	
\end{align}
We observe that, while the noise threshold for $N=50$ does not yet follow the asympotic scaling, the transition into the asymptotic scaling behavior is almost completed at $N=100$. With the noise threshold values obtained from the spline interpolations [Fig.~\ref{Fig:CAC_noise_threshold_scaling}(b)], we obtain the scaling fitting parameters $a \approx 0.46$ and $b \approx -1.07$. Similarly, the noise threshold values otained from the sigmoid fits deliver the scaling fitting parameters $a \approx 0.38$ and $b \approx -1.02$. If we take the averages of these values, and use their discrepancy to estimate the uncertainty, we obtain
\begin{align}
	a = 0.42 \pm 0.04 \hspace{5mm} \mathrm{and} \hspace{5mm} b = -1.04 \pm 0.02 .	
\end{align}
\end{subequations}
These scaling fitting parameters thus suggest that the noise threshold roughly follows a $1/N$ scaling behavior. Note that, to obtain these fitting parameters, we have used all available noise threshold values, including $N=50, 100$. If we exclude these from the fitting, then we obtain the scaling fitting parameters $a=0.52 \pm 0.08$ and $b=-1.08 \pm 0.04$, where we again averaged over the values obtained from the spline interpolations and the sigmoid fits.

The scaling law (\ref{Eq:CAC_noise_threshold_scaling_law}) allows us, for instance, to extrapolate at which problem size $N$ the noise threshold reaches system-intrinsic noise intensities, both in physical and in digital implementations. Below we will discuss examples for this.

We expect similar (polynomial) scaling behavior, possibly with modified fitting parameters, for other classes of Ising problems, for other noise models, and for other members of the CAC family $(\ref{Eq:classical_CIM})$ and $(\ref{Eq:CAC_error_variables_dynamics})$. In the following, we demonstrate this for sparse interaction matrices and for non-white noise models. More generally, we conjecture that similar scaling laws hold in general for deterministic analog combinatorial optimization solvers, as we will demonstrate below for an analog $k$-SAT solver.

\subsection{Sparse interaction matrices}

In order to clarify the noise threshold dependence on the structure of the interaction matrices $W_{ij}$, we modified the interaction model (\ref{Eq:dense_discrete_Ising_interaction}) towards allowing the possibility of vanishing interaction between two spins,
\begin{align} \label{Eq:sparse_discrete_Ising_interaction}
	W_{ij} \in \{-1,0,1\} ,
\end{align}
where the probability for a vanishing interaction matrix element, $W_{ij}=0$, is set to be $50\%$ (sparsity $50\%$).

We follow the same data generation and analysis procedures as above, restricting us to the problem sizes $N=50, 100, 150, 200$. The resulting noise threshold scaling is shown in Fig.~\ref{Fig:CAC_noise_threshold_scaling}(e), together with the noise threshold scaling of the dense interaction model.

We find that the noise threshold maintains a polynomial scaling behavior. Specifically, we obtain a best polynomial fit of the form
\begin{subequations} \label{Eq:CAC_sparse_noise_threshold_scaling_law}
\begin{align}
	D_{\rm th}(N) = a N^b ,
\end{align}
with $a \approx 0.54$ and $b \approx -1.20$ (spline interpolation) or $a \approx 0.46$ and $b \approx -1.17$ (sigmoid fit), resulting in the estimate
\begin{align}
	a = 0.50 \pm 0.04 \hspace{5mm} \mathrm{and} \hspace{5mm} b = -1.18 \pm 0.02 .
\end{align}
\end{subequations}
Note that, again, the problem size $N=50$ does not yet follow the asymptotic scaling behavior. In comparison with the dense interaction matrices, the scaling exponent $b$ is now marginally more negative, while the noise thresholds are shifted to smaller noise values. The latter may be surprising, as it indicates that sparse problems are more sensitive to the detrimental noise effect.

\subsection{Independence from noise model}

To demonstrate that the noise threshold scaling is robust under variations of the noise model, we now replace the white noise with colored noise. Recall that we use a constant time step $\Delta t = 2^{-6}$ to simulate the CAC equations with noise (\ref{Eq:CAC_with_noise}), in which case white noise is realized by applying statistically independent noise kicks at each time step. We now generalize this by introducing an effective correlation time $\tau > \Delta t$. This is achieved by applying the noise kicks at each time step with a probability $\prob \le 1$, which results in an average waiting time $\tau = \Delta t (1+\frac{1-\prob}{\prob})$ between noise kicks, where the limit of white noise corresponds to $\prob=1$. In effect, we complement the noise model (\ref{Eq:CAC_with_noise}) with a Poisson process $N(t) \in \{0,1\}$ with $\langle N(t) \rangle = {\rm prob}$:
\begin{subequations}
\begin{align}
	\sqrt{2 D} \xi_i^{(\sigma)}(t) &\rightarrow \sqrt{2 D} N_i^{(\sigma)}(t) \xi_i^{(\sigma)}(t) , \\
	\sqrt{2 D} \xi_i^{(e)}(t) &\rightarrow \sqrt{2 D} N_i^{(e)}(t) \xi_i^{(e)}(t) .
\end{align}
\end{subequations}
Note that we keep the (temporally uncorrelated) Gaussian noise processes independent of {\rm prob}, $\xi_i^{(\sigma)}(t) = \mathcal{N}(0, \Delta t^{-1})$ and $\xi_i^{(e)}(t) = \mathcal{N}(0, \Delta t^{-1})$.

In Fig.~\ref{Fig:CAC_noise_threshold_scaling}(f) we show, with problem size $N=100$ and for the four choices $\prob = 0.2, 0.5, 0.7, 1.0$, how the emergence of noise thresholds depends on the choice of $\prob$. Unsurprisingly, we find that the noise threshold is shifted to larger noise values with decreasing $\prob$. This is readily understood since the same noise strength $D$ per noise kick, in combination with a larger correlation time (i.e., a larger average waiting time between noise kicks), results in an overall reduced noise effect.

However, if we consider the effective noise intensity $\tilde{D} = D \times \prob$, we find that the resulting rescaled noise thresholds are invariant under variation of $\prob$, as shown in the inset of Fig.~\ref{Fig:CAC_noise_threshold_scaling}(f). This indicates that the emerging noise thresholds are robust under variations of the noise model, at least within the limits of the considered values of $\prob$. We expect that this holds for other modifications of our noise model, too. Irrespectively, it is clear that this invariance must break for sufficiently large $\tau$, for instance, if $\tau$ becomes comparable to the TTS.

 \section{Analog $k$-SAT solver} \label{Sec:Analog_k-SAT_solver}

As a second example, we analyze a class of dynamical systems that has been designed to solve Boolean satisfiability problems \cite{Ercsey2011optimization, Ercsey2012chaos, Molnar2013asymmetric, Sumi2014robust, Molnar2018continuous, Molnar2020accelerating}. Boolean satisfiability aims at finding an assignment to all Boolean variables in a Boolean formula such that the formula is evaluated as TRUE (assuming that such an assignment exists). Similar to the Ising problem, Boolean satisfiability is known to be, in general, NP-hard (even NP-complete), and has great practical relevance due to its relation to many decision, scheduling, and computational applications. There is strong evidence that the considered time-continuous dynamical solver \cite{Ercsey2011optimization} has, in principle, polynomial complexity in continuous time, however, at the expense of an exponential growth of the energy cost.

\subsection{Boolean satisfiability problems}

Boolean satisfiability, or $k$-SAT in short, assumes a Boolean formula composed of the conjunction (logical AND, denoted as $\land$) of $M$ clauses $c_m$, where each clause consists of the disjunction (logical OR, denoted as $\lor$) of $k$ Boolean variables $x_n$ or their negation $\overline{x}_n$, taken from a pool of $N$ variables. Boolean formulas that have this form are said to be in conjunctive normal form (CNF). Boolean variables are binary, taking only the values TRUE or FALSE, and a solution consists of an assignment of truth values to all variables such that the formula evaluates as TRUE, that is, the assignment satisfies all clauses. Boolean satisfiability, $k$-SAT, is restricted to formulas that admit at least one satisfying assignment of truth values. The more general situation, where such an assignment does not necessarily exist, and where instead the number of satisfied clauses is to be maximized, constitutes a different problem class termed MAX-SAT, which we do not consider here.

Boolean satisfiability is known to be NP-hard for any $k\ge3$. Following \cite{Ercsey2011optimization}, we restrict ourselves here to the case $k=3$, i.e., the 3-SAT problem class. A 3-SAT formula in CNF with $N$ variables and $M$ clauses takes the form
\begin{subequations}
\begin{align}
	f = c_1 \land c_2 \land \dots c_M ,
\end{align}
where each clause is the disjuntion of 3 variables,
\begin{align}
	c_m = (X_{m,1} \lor X_{m,2} \lor X_{m,3}) ,
\end{align}
\end{subequations}
with $X_{m,i} \in \{ x_1, \overline{x}_1, x_2, \overline{x}_2, \dots, x_N, \overline{x}_N \}$. The ratio $\alpha=M/N$ between the number of clauses and the number of variables has a strong impact on the problem complexity: Satisfiable assignments are typically easy to find for $\alpha < 4.21$, while problems with $\alpha > 4.26$ are in most cases easily shown not to be satisfiable; hard problems with exponentially long search times are usually located in the ``frozen'' regime around $\alpha \approx 4.25$. In our numerical analysis below we exclusively focus on 3-SAT problems with $\alpha = 4.25$.

\subsection{Dynamical system}

We now introduce the analog $k$-SAT solver in the form of a system of coupled differential equations \cite{Ercsey2011optimization}. To this end, we first explain how the clauses of a given $k$-SAT formula can be expressed in terms of constraint functions that play the role of the constituents of a potential function. For a Boolean formula in CNF form with $N$ variables and $M$ clauses, we introduce continuous variables $s_i \in [-1,1]$ such that $s_i = -1$ corresponds to $x_i =$ FALSE and $s_i = 1$ corresponds to $x_i =$ TRUE. A Boolean formula is then represented as a matrix $c$ of dimension $M \times N$ such that $c_{mi} = 0$ if $s_i$ is absent in the $m$th clause, $c_{mi} = 1$ if the $i$th variable is present in the natural form ($x_i$), and $c_{mi} = -1$ if the $i$th variable is present in the $m$th clause in the negated form ($\overline{x}_i$). With this we can define a constraint function $K_m$ for each clause $c_m$:
\begin{align}
	K_m(\vec{s}) = 2^{-k} \prod_{i =1}^{N} (1-c_{mi}s_i) .
\end{align}
A modified energy function is now defined in terms of the constraint functions as
\begin{align}
	V(\vec{s},\vec{a}) = \sum_{m=1}^{M} a_m K_m(\vec{s})^2 ,
\end{align}
where we complement each constraint function with an auxiliary variable $a_m \in (0,\infty)$ that decisively supports the solution finding, as explained below. Note that $V(\vec{s},\vec{a})=0$ indicates a solution $\vec{s}=\vec{s}^*$ of the $k$-SAT problem.

The core of the solution-finding mechanism is a gradient descent towards the global minimum $V(\vec{s},\vec{a})=0$ of the potential landscape. To avoid the trapping in local non-solution minima, the dynamics of the auxiliary variables are designed to amplify the energy contributions of unsatisfied constraints, which effectively destabilizes local minima. This evolution is captured by the following continuous-time deterministic dynamical system:
\begin{subequations} \label{Eq:kSAT_dynamical_system}
\begin{align}
	\frac{ds_i}{dt} &= (-\nabla_s V(\vec{s},\vec{a}))_i \nonumber \\
	 				&= \sum_{m=1}^{M} 2 a_m c_{mi} K_{mi}(\vec{s}) K_{m}(\vec{s}) , \hspace{2mm} i = 1, \dots, N \\
	\frac{da_m}{dt} &= a_m K_m(\vec{s}) , \hspace{2mm} m = 1, \dots, M ,
\end{align}
\end{subequations}
where $\nabla_s$ is the gradient operator with respect to $s$ and $K_{mi} = K_{m}/(1-c_{mi}s_i)$. Note that, with positive initial conditions $a_m(0)>0 \hspace{1mm} \forall m$, the auxiliary variables monotonically grow unless their respective constraint is satisfied, in which case they remain constant. The main variables $s_i$,  on the other hand, can be shown to be confined to the interval $[-1,1]$, towards which they rapidly converge in case the initial conditions lie outside.

The dynamical system (\ref{Eq:kSAT_dynamical_system}) has several notable properties. Foremost, it is guaranteed to converge to $k$-SAT solutions $\vec{s}^*$, which are by construction the only fixed points of the dynamics, excluding also the existence of limit cycles. Moreover, there is strong evidence that solutions are generically found in polynomial continuous time; however at the cost of, in general, exponentially growing energy fluctuations dictated by the auxiliary variables, which, in turn, can be seen as exponentially accelerating the dynamics with time. Note that variables $s_i$ that are undetermined by the constraints may converge to values other than $\pm1$ but within $[-1,1]$.

The transient dynamics of the main variables $s_i$, i.e., the dynamics before the system has converged to its fixed point, are generically chaotic for hard problems, expressed by rapid fluctuations of the variables and an exponential sensitivity to the initial conditions. While this chaotic stage is essential in that it implements the solution search, it is clear that the dynamics, which explore increasingly small length scales with growing problem size, become potentially highly susceptible to noise perturbations. While this may suggest a weak robustness against noise, the strong convergence towards the fixed points, on the other hand, which holds before and after perturbations alike, may rather speak in favor of a distinguished noise robustness. Below we explore how these two competing tendencies play out.

While both the CAC algorithm (\ref{Eq:classical_CIM}, \ref{Eq:CAC_error_variables_dynamics}) and the $k$-SAT solver (\ref{Eq:kSAT_dynamical_system}) display chaotic dynamics, the latter is distinguished in that the chaotic dynamics is transient and the system will eventually converge towards the solution as a stationary state, which can serve as a halting condition. Moreover, in digital implementations the CAC algorithm is typically run with a constant step size (e.g., in efficient FPGA deployments), while the $k$-SAT solver must rely on integrators with an adaptive step size to accurately reproduce the chaotic dynamics. Strong numerical evidence \cite{Ercsey2011optimization} suggests that the step size reduction required by the chaotic dynamics results in an overall exponential run time scaling with problem size, in effect neutralizing the polynomial scaling with respect to the continuous time of the dynamical model.

\subsection{Noise model}

The effect of noise on the dynamical system (\ref{Eq:kSAT_dynamical_system}) was previously investigated in~\cite{Sumi2014robust}. The authors demonstrate, for problem sizes ranging from $N=20$ to $N=100$, assuming Gaussian white and colored (Ornstein-Uhlenbeck process) noise with small to moderate noise intensities, and using an Euler-Maruyama integrator with a fixed step size, that the dynamical system (\ref{Eq:kSAT_dynamical_system}) is, under these conditions, remarkably robust against noise. Specifically, the authors' goal was to demonstrate, for the considered problem sizes, the viability of implementing the $k$-SAT solver (\ref{Eq:kSAT_dynamical_system}) on real hardware, e.g., analog circuits.

Our approach here is different in significant ways. Most importantly, we are interested in the maximum noise acceptable before the solution-finding capability breaks down, and how this noise threshold scales with the problem size. To this end, we deliberately ramp up the noise intensity to sufficiently large levels to reach the noise threshold, irrespective of whether these noise levels are realistic or not in specific analog hardware. The rationale in this approach is to leverage the extracted scaling behavior to extrapolate the expected noise response to problem sizes that are beyond the scope of our numerical analysis.

Secondly, our noise model is realized through temporally equidistant noise kicks. While this corresponds to a specific colored noise model, we also expect here, as it is the case with the CAC algorithm, that the noise response is generic and widely independent of the specifics of the noise model. A core advantage of this noise model is that, between noise kicks, we can conveniently use an ordinary differential equation (ODE) integrator with adaptive step size (we use the RK45 method). This way we avoid introducing an additional intrinsic noise source due to, e.g., a finite (constant) step size. Considering that chaotic dynamics generically cover a broad range of length scales (which is expected to further grow with the problem size, in particular towards smaller scales), our approach maintains a faithful expression of the dynamical system (\ref{Eq:kSAT_dynamical_system}) in between the noise kicks.

The dynamical system (\ref{Eq:kSAT_dynamical_system}), complemented by the noise model, can be written as
\begin{subequations} \label{Eq:kSAT_solver_with_noise}
\begin{align}
	\frac{ds_i}{dt} &= \sum_{m=1}^{M} 2 a_m c_{mi} K_{mi}(\vec{s}) K_{m}(\vec{s}) + \sqrt{2D} \zeta_i(t) \delta(t {\rm mod} \Delta t) \label{Eq:kSAT_system_variable_dynamics_with_noise} \\
	\frac{da_m}{dt} &= a_m K_m(\vec{s}) \label{Eq:kSAT_auxiliary_variable_dynamics} ,
\end{align}
\end{subequations}
where $\Delta t$ stands for the temporal separation of the noise kicks, $D$ scales the kick strength, and $\zeta_i(t) = \mathcal{N}(0, \Delta t^{-1})$ denotes a temporally uncorrelated (more precisely, $\tau \ll \Delta t$, with $\tau$ the correlation time) normal random variable with vanishing mean and variance $1/\Delta t$. Throughout our analysis, we choose $\Delta t=0.5$, which lies well below the (noise-free) average time-to-solution for the smallest problem size considered, $N=40$.

We point out that we assume the auxiliary variables $a_m(t)$ to be noise free. This can be justified conceptually, as the auxiliary variables can be easily eliminated by inserting the formal solution of (\ref{Eq:kSAT_auxiliary_variable_dynamics}) into (\ref{Eq:kSAT_system_variable_dynamics_with_noise}). From a more practical perspective, this eliminates the possibility that the noise drives some auxiliary variables into negative values, a situation where the dynamical system loses its solution-finding capacity, unless the affected auxiliary variables recover into positive values before they run away exponentially fast into the negative. While this could be easily fixed by restricting the auxiliary variables to positive values (for instance, by reflecting the noise kicks at the origin), the generically growing nature of the auxiliary variables and their specific role in the system dynamics (\ref{Eq:kSAT_system_variable_dynamics_with_noise}) (namely, to weigh the constraint functions relative to each other) ultimately suggest that noise in the auxiliary variables does not significantly affect the solution-finding process.

\subsection{Data generation}

We generate, for each problem size $N$, 5000 random 3-SAT problem instances with $\alpha=4.25$. Recall that, by definition, all these problem instances are satisfiable. Each problem instance is evaluated for an $N$-dependent range of kick strengths $D$. The initial conditions for the system variables are fixed as $s_i(0)=0, i \in \{1, \dots, N\}$, while the initial conditions of the auxiliary variables are fixed as $a_m(0)=1, m \in \{1, \dots, M\}$. The problem sizes span one order of magnitude, from $N=40$ to $N=400$, from $N=40$ to $N=100$ in steps of 20, and from $N=100$ to $N=400$ in steps of 50.

\begin{figure*}[htb]
	\centering
	\includegraphics[width=0.99\textwidth]{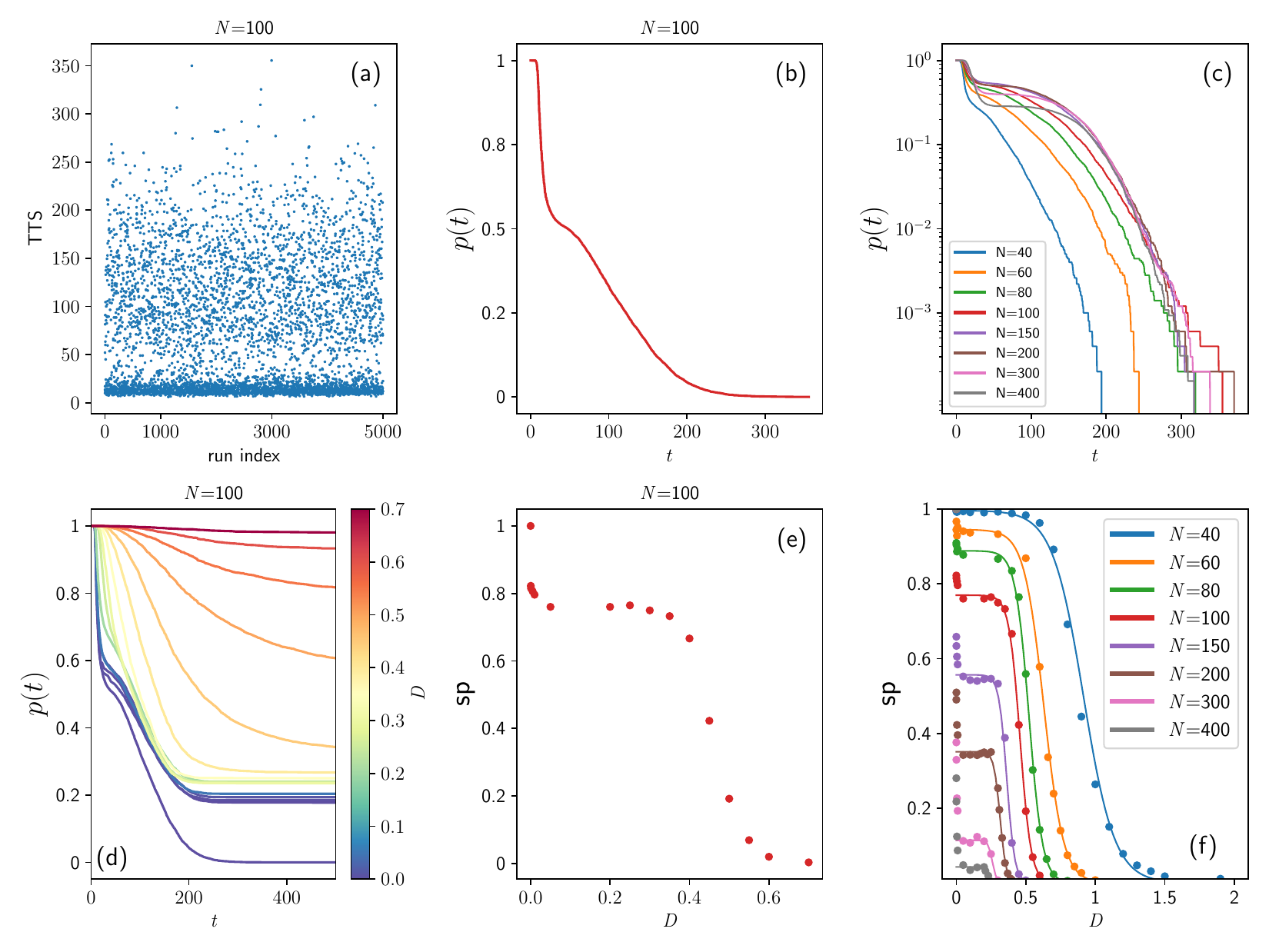}
	\caption{\label{Fig:kSAT_data_processing} Data generation and processing for the $k$-SAT solver (\ref{Eq:kSAT_dynamical_system}), and the emergence of noise thresholds. (a) Time-to-solution (TTS) data, obtained with the dynamical system (\ref{Eq:kSAT_dynamical_system}) in the absence of noise, for 5000 random $3$-SAT instances of problem size $N=100$ and clause-variable ratio $\alpha=4.25$. The initial conditions for both the spin variables and the auxiliary variables are fixed throughout (see main text). If the algorithm does not find a problem-solving sign configuration of the spin variables before the cutoff time $T_{\rm cut}=N^2=10^4$, $T_{\rm cut}$ is assigned as TTS. While this does not occur in the noise-free case, it becomes increasingly relevant in the presence of noise. (b) Fraction of unsolved problems $p(t)$ for the same data. By definition, $p(t)$ monotonically decreases from $p(0)=1$. Moreover, $p(t)$ decays to zero, indicating that all problem instances are solved within the considered time range. (c) Fraction of unsolved problems $p(t)$ for all problem sizes from $N=40$ to $N=400$. With increasing $N$ the $p(t)$ flattens, reflecting problem hardness on average. (d) Fraction of unsolved problems $p(t)$ for $N=100$ in the presence of noise, cf.~Eq.~(\ref{Eq:kSAT_solver_with_noise}). For sufficiently large noise kick strengths $D$, $p(t)$ assumes constant asymptotic values, implying that the corresponding fraction of problems becomes unsolvable. (e) Success probability ${\rm sp}(D)$ as a function of the noise kick strength $D$. The emergence of a noise threshold, where noise prohibits any solution finding, is preceded by the filtering of a constant fraction of problem instances with onset at small noise values. We refer to the former as the hard noise threshold, and to the latter as the soft noise threshold. (f) Success probabilities ${\rm sp}(D)$ for all problem sizes from $N=40$ to $N=400$. The hard noise threshold moves to smaller noise values with increasing problem size. At the same time, the soft noise threshold remains approximately independent of the problem size, while the fraction of filtered problem instances increases rapidly with $N$. The data points are, after the onset of the soft noise threshold, fit with a sigmoid function (see main text).}
\end{figure*}

We consider a run successful as soon as the system variables $s_i(t)$ assume a sign configuration that satisfies the 3-SAT formula. The corresponding time $t$ is then recorded as the time-to-solution (TTS) of the run. In the absence of noise this happens before $t=500$ for all considered problem sizes. In the presence of noise we wait until $T_{\rm cutoff} = N^2$ to see if a solution is found, and if no solution is found before $T_{\rm cutoff}$ is reached, we assign $T_{\rm cutoff}$ as TTS, which is equivalent to labeling the run as failed. In Fig.~\ref{Fig:kSAT_data_processing}(a) we show the noise-free TTS distribution for $N=100$.

Similar to the CAC data processing, the next step is to derive the fraction of unsolved problems, defined as in (\ref{Eq:fraction_of_unsolved_problems}), where $N_{\rm sol}(t)$ now represents the number of instances solved before $t$, and $N_{\rm run} = 5 \times 10^3$. In Fig.~\ref{Fig:kSAT_data_processing}(b) we show the fraction of unsolved problems $p(t)$ for $N=100$ [the TTS data depicted in Fig.~\ref{Fig:kSAT_data_processing}(a)]. We observe that, after an initial rapid decay of $p(t)$, it intermediately flattens, indicating that there is a time window whitin which only few problems are solved, before it assumes the asymptotic exponential decay that is also described in \cite{Ercsey2011optimization}.

The fractions of unsolved problems in the absence of noise for all considered problem sizes are shown in Fig.~\ref{Fig:kSAT_data_processing}(c). Unsurprisingly, we observe again (as for the case of the CAC algorithm) that the decay of $p(t)$ flattens out with increasing $N$, indicating increasing problem hardness. Note that all problem sizes exhibit a transitional stage before reaching the asymptotic exponential decay regime. Moreover, deviations from the exponential decay behavior when approaching the maximal TTS within the sample, ${\rm TTS}_{\rm max}$, necessarily occur due to our finite sample sizes and since, by definition, $p({\rm TTS}_{\rm max})=0$.

The effect of noise is demonstrated in Fig.~\ref{Fig:kSAT_data_processing}(d) for $N=100$ and a span of noise kick strengths $D$. We observe that moderate kick strengths still admit the majority of problem instances to be solved, while a constant, persistent fraction of problem instances remains unsolved for arbitrary times (as indicated by the flattened out $p(t)$). Suffciently large kicks eventually prevent all instances to converge to a solution, indicating the emergence of a noise threshold. We emphasize that, while this is not obvious from Fig.~\ref{Fig:kSAT_data_processing}(d), all p(t) converge to finite asymptotic values in the presence of noise.

\subsection{Hard-noise threshold scaling}

Similar to the discussion for the CAC algorithm in the previous section, we introduce the success probability ${\rm sp}(D)$, that is, the fraction of problems that remains unsolved no matter how long we wait, as a measure of the noise impact. In contrast to (\ref{Eq:CAC_success_probability}), there is no need here to renormalize, as all problems are solvable in the absence of noise if $T_{\rm cut}$ is chosen sufficiently large. We thus define
\begin{align}
	{\rm sp}(D) = 1 - p_D(T_{\rm cut}) .
\end{align}
In Figure~\ref{Fig:kSAT_data_processing}(e) we show ${\rm sp}(D)$ for our data with $N=100$. Interestingly, we observe a dropoff of ${\rm sp}(D)$ in two stages: First, at small noise kick strength $D \approx 0.05$ the success probability drops to a plateau value of ${\rm sp}(0.4 \gtrsim D \gtrsim 0.05) \approx 0.75$, indicating the early noise-induced filtering of a fraction of about $25\%$ of problem instances. This early dropoff of the success probability has also been reported in~\cite{Sumi2014robust}. We refer to this partial noise threshold as the {\it soft noise threshold}.

The second and complete dropoff of the success probability occurs at around $D\approx 0.5$, indicating that beyond this point all problem instances become unsolvable. We refer to this threshold as the {\it hard noise threshold}. The hard noise threshold corresponds to the CAC noise threshold reported in the previous section and can generically be expected to emerge in all deterministic (or probabilistic) analog combinatorial optimization algorithms.

Similar soft and hard noise thresholds emerge for other problem sizes $N$, as shown in Fig.~\ref{Fig:kSAT_data_processing}(f). We find that the hard noise thresholds consistently decrease with increasing $N$, together with the widths of the corresponding transitions and in agreement with the pattern observed for the CAC algorithm. At the same time, we find, in agreement with~\cite{Sumi2014robust}, that the soft noise thresholds appear to be basically independent of the problem size, while the fraction of filtered problems increases monotonically with increasing problem size. We will analyze the scaling of the filtered fraction at the soft noise threshold in the next subsection.

To quantify the hard-noise threshold behavior, we use sigmoid functions ${\rm sp}(D) = {\rm sp}_{\rm sth}/[1-{\rm exp}(-(D-D_{\rm th})/\Delta D)]$ to fit the success probabilities for noise values after passage of the soft noise threshold, that is, after the success probabilities have dropped to their intermediate plateau values ${\rm sp}_{\rm sth}$, cf.~Fig.~\ref{Fig:kSAT_data_processing}(f). From these fits we then extract the hard-noise threshold values $D_{\rm th}$ and the associated transition widths $\Delta D$ (the latter are not analyzed further).

\begin{figure}[htb]
	\includegraphics[width=0.99\columnwidth]{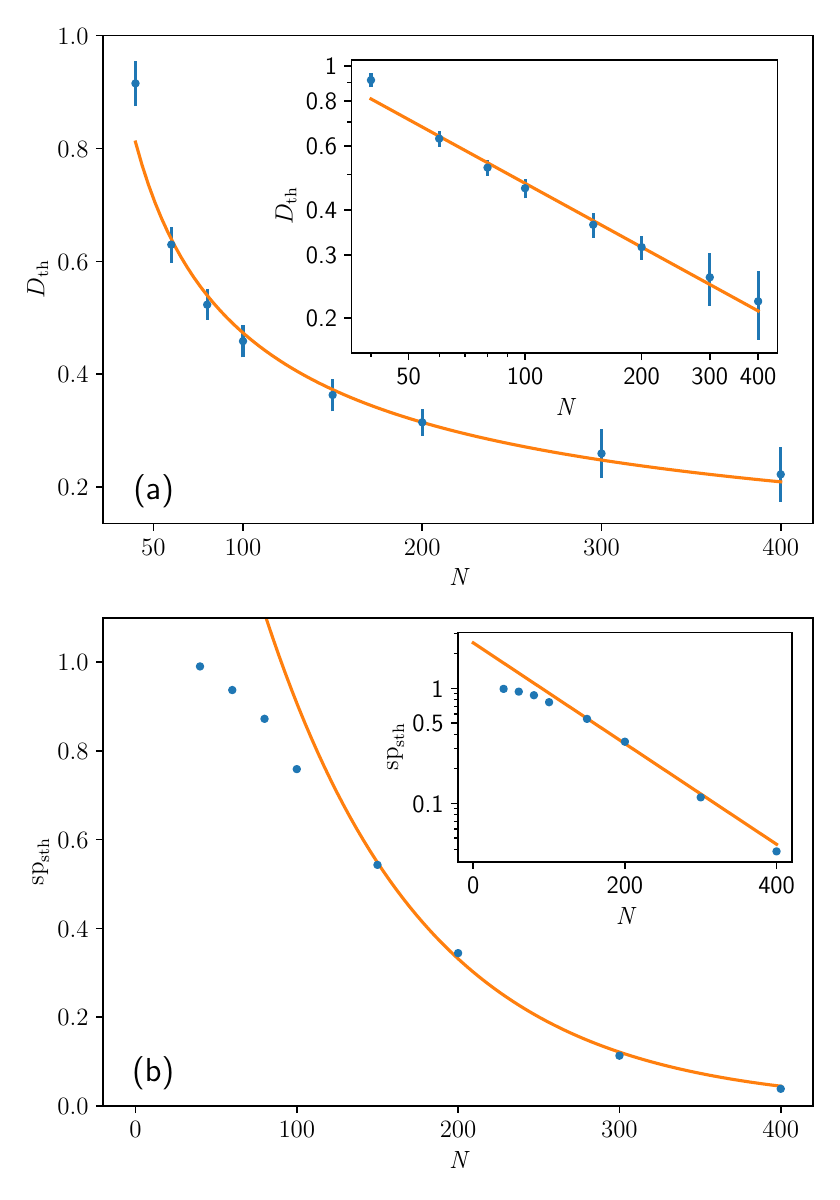}
	\caption{\label{Fig:kSAT_noise_thresholds_scaling} Hard and soft noise threshold for the $k$-SAT solver (\ref{Eq:kSAT_dynamical_system}). (a) Scaling of the hard noise threshold with problem size $N$. The extracted threshold values (blue dots) follow a polynomial asymptotic scaling of the form $a N^b$ (orange solid line), with $a\approx  7.10$ and $b \approx -0.59$, cf.~(\ref{Eq:kSAT_hard_noise_threshold_scaling}). The inset shows the same data in double-logarithmic representation. Except for $N=40$, all problem sizes have converged to the asymptotic scaling behavior. (b) Scaling with $N$ of the success probability ${\rm sp}_{\rm sth}$ at the intermediate plateau assumed after passage of the soft noise threshold. The extracted plateau success probabilities (blue dots) follow asymptotically an exponential scaling (orange solid line), see also~(\ref{Eq:kSAT_soft_noise_threshold_scaling}). The inset depicts the same data in logarithmic representation.}
\end{figure}

The scaling of the hard noise threshold with the problem size $N$ is shown in~Fig.~\ref{Fig:kSAT_noise_thresholds_scaling}(a). We find that the hard noise thresholds for all examined problems sizes, except for a slight deviation for the smallest problem size ($N=40$), which we interpret as an incomplete convergence to the asymptotic scaling behavior, display a regular polynomial behavior. Indeed, we get excellent agreement with a polynomial fit of the form
\begin{subequations} \label{Eq:kSAT_hard_noise_threshold_scaling}
\begin{align}
	D_{\rm th}(N) = a N^b ,
\end{align}
as indicated by the solid line in Fig.~\ref{Fig:kSAT_noise_thresholds_scaling}(a). The corresponding fitting parameters are
\begin{align}
	a = 7.10 \pm 0.09 \hspace{5mm} \mathrm{and} \hspace{5mm} b = -0.59 \pm 0.02 .
\end{align}
\end{subequations}
The fitting exponent $b$ thus suggests that the hard noise threshold roughly follows a $1/\sqrt{N}$ scaling behavior, which appears milder than the rough $1/N$ scaling behavior that we obtained for the CAC algorithm. While this may speak in favor of the noise robustness of the $k$-SAT solver, we will see in the next subsection that the soft noise threshold introduces additional challenges that are absent for the CAC algorithm.

It is worth noting that the noise kick strengths required to reach the hard-noise threshold regime are, especially for the smaller problem sizes, strong enough to constantly kick the system variables out of their regular range of $[-1,1]$. While this may demonstrate the intrinsic self-healing properties of the dynamical system (\ref{Eq:kSAT_dynamical_system}), such strong noise may appear unrealistic from a practical perspective. However, let us recapitulate that the main goal of our analysis is to leverage the extracted scaling behavior to infer the impact of realistic noise levels at problem sizes that are challenging to investigate numerically.

\subsection{Soft-noise threshold scaling}

While the success probability assumes for all considered problem sizes an intermediate plateau value at about the same noise strength $D_{\rm sth} \approx 0.05$, which we denote the soft noise threshold, the plateau values ${\rm sp}_{\rm sth}$ strongly depend on the problem size, as shown in~Fig.~\ref{Fig:kSAT_noise_thresholds_scaling}(b). Remarkably, we find good asymptotic agreement with an exponential fitting function of the form
\begin{subequations} \label{Eq:kSAT_soft_noise_threshold_scaling}
\begin{align}
	{\rm sp}_{\rm sth}(N) = a e^{b N} .
\end{align}
The corresponding fitting parameters are
\begin{align}
	a = 2.5 \pm 0.2 \hspace{5mm} \mathrm{and} \hspace{5mm} b = -0.0101 \pm 0.0005 .
\end{align}
\end{subequations}
This suggests that the soft noise threshold rapidly becomes a relevant factor of the noise response and, indeed, eventually forestalls the hard threshold by effectively filtering all problem instances. The question, beyond which problem size the hard threshold comes to effect before the soft threshold and thus replaces the latter, is discussed in the next section.

The occurrence of an $N$-independent soft noise threshold has also been reported in \cite{Sumi2014robust}, for white and (Ornstein-Uhlenbeck type) correlated noise. Specifically, the authors show for problem sizes $N=30, 50, 100$ that the noise impact is negligible for noise intensities up to $I=1.5\times 10^{-3}$. In our noise model, this approximately corresponds to noise kick strengths $D\approx I/\Delta t^2 = 3.75 \times 10^{-5}$, which is too small to be resolved by our analysis. Note, however, that our soft-noise threshold value indicates the {\it completion} of the dropoff to the intermediate plateau value, while the value $I=1.5\times 10^{-3}$ indicates the {\it onset} of the dropoff.

The authors of \cite{Sumi2014robust} obtain, for $N=100$, $\alpha=4.25$, and white noise, a drop of the success probability below $0.4$ at $I=5 \times 10^{-3}$ (their largest noise value, which is still before the plateau is reached). Comparing this to our ${\rm sp}_{\rm sth}(N=100) \approx 0.75$, we infer that the noise impact of the white noise model of \cite{Sumi2014robust} is more severe than the noise impact of our noise model (\ref{Eq:kSAT_solver_with_noise}). This discrepancy by about a factor 2 may, at least in part, be explained by the different noise treatment of the auxiliary variables, as we assume the auxiliary variables to be noisefree, while they are exposed to noise in \cite{Sumi2014robust}. Similar behavior is found in \cite{Sumi2014robust} for finite correlation time $\tau=1$.  The emergence of the intermediate plateau and its scaling with $N$, and the emergence of a hard threshold and its scaling with $N$, are not investigated in \cite{Sumi2014robust}.

\section{Discussion} \label{Sec:Discussion}

In the following, we will use the scaling laws that we extracted in the previous sections to predict noise-induced limitations on the applicability of the two considered analog solvers, with regard to both digital implementations and actual physical realizations. Specifically, we consider the following three cases: For the Ising solver, we analyze (i) limitations on the problem size $N$ in typical FPGA-based digital implementations, and (ii) the required amplitude scaling with $N$ in CIM (Coherent Ising Machine) realizations to maintain a viable signal-to-noise-ratio due to fundamental quantum noise. For the $k$-SAT solver, we determine the problem size $N$, at which the hard noise threshold drops below the soft noise threshold and thus replaces the latter as the limiting noise threshold.

\subsection{Digital implementation of the Ising solver}

Digital implementations of analog dynamical systems provide promising avenues for their deployment. This holds in particular for FPGAs (Field-Programmable Gate Arrays), which can be tailored and optimized for such special-purpose applications. One way to achieve tangible speed-up with FPGAs relies on fixed-point number encoding, which, however, imposes a lower bound on number precision, as variables are confined to an equidistant number grid. The true variable value $x$ is then decomposed into its nearest grid point $\tilde{x}$ and a random variable uniformly drawn from $\xi \in [-\Delta x/2, \Delta x/2]$, $x = \tilde{x} + \xi$, where $\Delta x$ denotes the distance between two adjacent points on the equidistant number grid. The resulting variance $\sigma_x^2 = \Delta x^2/12$ quantifies the average deviation of the grid point $\tilde{x}$ from the true value $x$. This shows that the grid spacing $\Delta x$ can be interpreted as an induced noise scale. In the following, we  estimate the maximum solvable problem size that results from this inherent noise source.

In order to see how the resolution-induced noise impacts the dynamics, we first note that, for sufficiently large $N$, the dominant noise contribution to $d x_i/d t$ in (\ref{Eq:classical_CIM}) stems from the term $\epsilon e_i \sum_{j} W_{ij} x_j$. For noisy variables $\tilde{x}_j$, its noise contribution is described by a random variable $Z_i = \tilde{e}_i \sum_{j} W_{ij} \xi_j$, where $\tilde{e}_i$ denotes the nearest grid representation for $\epsilon e_i$. Using the central limit theorem, we can assume that, for sufficiently large $N$, $Z_i$ is described by a normal distribution with vanishing mean, $\langle Z_i \rangle = 0$, and variance $\sigma_{Z, i}^2 = \tilde{e}_i^2 \sigma_x^2 N$. Note that, with (\ref{Eq:dense_discrete_Ising_interaction}), the $W_{ij}$ randomly change the signs of the $x_j$ and thus can be absorbed by the random variables $\xi_j$.

If we match the variance $\sigma_{Z}^2 = \tilde{e}^2 \sigma_x^2 N$ (note that we replaced $\tilde{e}_i$ by a typical $\tilde{e}$) with the variance of our noise model (\ref{Eq:CAC_with_noise}), $\tilde{e}^2 \sigma_x^2 N = 2 D/\Delta t$, we can express the noise intensity as
\begin{align}
	D = \frac{1}{24} \tilde{e}^2 \Delta t \Delta x^2 N ,
\end{align}
where we used $\sigma_x^2 = \Delta x^2/12$. If we match this with the threshold scaling law (\ref{Eq:CAC_noise_threshold_scaling_law}), $a N^b = \tilde{e}^2 \Delta t \Delta x^2 N/24$, we can solve for the maximal problem size that can be treated successfully with the given parameters,
\begin{align} \label{Eq:CAC_precision_scaling}
	N_{\rm max} = \left( \frac{\tilde{e}^2 \Delta t \Delta x^2}{24 a} \right)^{1/(b-1)} .
\end{align}

In order to evaluate (\ref{Eq:CAC_precision_scaling}), we assume that the fixed-point number encoding in a FPGA implementation is based on 18 bits, where 1 bit is reserved for the sign, 5 bits for the integer, and 12 bits for the decimal. This results in a grid spacing $\Delta x = 2^{-12}$. Moreover, we assume, for a speedy convergence, a rather coarse time step $\Delta t = 1/2$. Finally, we make the generic assumption that $\tilde{e} \approx 1$. With the scaling paramters $a = 0.42$ and $b = -1.04$ [cf.~(\ref{Eq:CAC_noise_threshold_scaling_law})], we can then estimate that $N_{\rm max} \approx 15000$. Note that, if we approximate $b \approx -1$, the maximal problem size roughly scales as $N_{\rm max} \propto 1/(\sqrt{\Delta t} \Delta x)$.

Let us clarify that the scaling of the maximal problem size with the available number precision, Eq.~(\ref{Eq:CAC_precision_scaling}), is not restricted to digital implementations of the Ising solver. More generally, it applies whenever the resolution of the variables is limited. Conversely, the required variable precision scales with the problem size as $\Delta x \propto N^{(b-1)/2}$, or, with $b \approx -1$, $\Delta x \propto 1/N$. This can be reformulated as resource scaling, in the sense that the number of digits (per variable) that are actively involved in the calculation scales as $N$.

\subsection{Signal-to-noise-ratio for quantum noise}

As explained in Sec.~\ref{Sec:Analog_Ising_solver}, the CAC algorithm has been inspired by a specific physical computing platform based on degenerate optical parametric oscillators (DOPOs) \cite{Wang2013coherent, Marandi2014network, Mcmahon2016fully, Inagaki2016coherent, Yamamoto2017coherent, Yamamoto2020coherent, Honjo2021spin, Zhou2024frustration}. While the CAC algorithm has not been implemented in Coherent Ising Machines (CIMs), yet, this is in principle conceivable in a measurement-feedback scheme, where the dynamics of the error variables are calculated separately on a digital computer.

In CIMs, the coherent-state amplitude $\alpha$ of the DOPO takes the role of the soft spin variable, while the Heisenberg uncertainty $\sigma_q$ sets a fundamental limit to the resolution. The inverse signal-to-noise-ratio $\sigma_q/|\alpha|$ then corresponds to the precision $\Delta x$ in the previous subsection, $\Delta x = \sigma_q/|\alpha|$. It follows immediately from the discussion in the previous subsection that, in order to maintain the required precision, the amplitude must scale as $|\alpha| \propto N^{(1-b)/2}$ (or $|\alpha| \propto N$ if $b \approx -1$). Since the amplitude scales with the pump $p$ as $|\alpha| \propto \sqrt{p}$, the pump should scale as $p \propto N^{1-b}$ (or $p \propto N^2$ if $b \approx -1$). This arguably sets an upper limit in terms of problem size to the implementation of the CAC algorithm with physical CIMs.

\subsection{Soft-noise threshold passage}

For the $k$-SAT solver, the exponential-in-$N$ drop-off of the success probability after passing the soft noise threshold at $D_{\rm sth} \approx 0.05$, Eq.~(\ref{Eq:kSAT_soft_noise_threshold_scaling}), renders it imperative to keep noise levels below the soft noise threshold already for intermediate problem sizes. Indeed, already for $N=1000$, the scaling law~(\ref{Eq:kSAT_soft_noise_threshold_scaling}) predicts that the success probability drops to ${\rm sp}_{\rm sth}(1000) = 10^{-4}$ if the soft noise threshold is surpassed, essentially filtering all problem instances. On the other hand, the hard noise threshold is predicted to remain above the soft noise threshold at $D_{\rm th}(1000) \approx 0.12$ according to (\ref{Eq:kSAT_hard_noise_threshold_scaling}).

This apparent anomaly, that the noise level can be kept constant with increasing problem size, ceases to apply for problem sizes where the hard noise threshold has dropped below the soft noise threshold. This passage of the soft noise threshold is predicted to occur at $N \approx 4500$, that is, $D_{\rm th}(4500) \approx D_{\rm sth}$ [cf.~(\ref{Eq:kSAT_hard_noise_threshold_scaling})]. In other words, for problem sizes larger than $N \approx 4500$ the noise level must follow the scaling (\ref{Eq:kSAT_hard_noise_threshold_scaling}) of the hard threshold so that the solver can deliver solutions.

\section{Conclusions} \label{Sec:Conclusions}

We investigated the effect of noise on the solution finding capability of two deterministic analog combinatorial optimization solvers, the Chaotic Amplitude Control (CAC) algorithm, which is tailored for Ising problems, and a $k$-SAT solving algorithm. We find that both solvers exhibit noise thresholds beyond which the solution-finding capability rapidly deteriorates. Moreover, these thresholds scale, for both solvers, polynomially with the problem size, as indicated by Eqs.~(\ref{Eq:CAC_noise_threshold_scaling_law}) (CAC) and (\ref{Eq:kSAT_hard_noise_threshold_scaling}) ($k$-SAT solver). As a special feature, the $k$-SAT solver exhibits an additional, problem size-independent soft noise threshold, which is relevant at small to intermediate problem sizes and beyond which the success probability drops exponentially with problem size, cf.~Eq.~(\ref{Eq:kSAT_soft_noise_threshold_scaling}).

We believe that the two investigated solvers are sufficiently diverse to provide a representative sample of the class of deterministic analog combinatorial optimization solvers. Therefore, we expect that other members of this class exhibit similar noise thresholds that scale polynomially with the problem size. We underpin this expectation by demonstrating that both dense and sparse Ising interactions give rise to (different) polynomially scaling noise thresholds for the CAC algorithm. Moreover, we demonstrate the robustness of the polynomial scaling behavior under white and colored noise models.

The uncovered noise threshold scaling laws allow us to deduce fundamental constraints on the acceptable noise levels in such deterministic analog combinatorial optimization solvers. This is similar to the error budget in near-term quantum computers without quantum error correction, which also operate as analog computers, where errors limit both the number of quantum bits that can be employed in a computation and the circuit depth.

On the other hand, we can use the scaling laws to derive, for given noise levels, the reliable application range of these solvers. Importantly, this includes all possible noise sources, ranging from errors due to finite variable precision to errors due to time discretization, and applies both to realizations with actual physical systems as well as digital emulations. For instance, we use the CAC noise threshold scaling to estimate the maximal problem size that can be addressed in typical FPGA emulations.

From a more generic perspective, we can interpret the deduced noise thresholds and their scalings in terms of resource costs that are inherent in deterministic analog combinatorial optimization solvers that harness transiently chaotic dynamics. For instance, conditions on variable precision directly translate into memory cost, as higher precision implies larger memory allocation for the storage of variables.

\begin{acknowledgments}
The computations in this work were performed using the HOKUSAI BigWaterfall system at the Information System Division of RIKEN.
C.G. is partially supported by RIKEN Incentive Research Projects.
F.N. is supported in part by:
Nippon Telegraph and Telephone Corporation (NTT) Research,
the Japan Science and Technology Agency (JST)
[via the CREST Quantum Frontiers program Grant No. JPMJCR24I2,
the Quantum Leap Flagship Program (Q-LEAP),
and the Moonshot R\&D Grant Number JPMJMS2061],
and the Office of Naval Research (ONR) Global (via Grant No. N62909-23-1-2074).
\end{acknowledgments}



\bibliography{noise_resilience}

\begin{thebibliography}{36}%
\makeatletter
\providecommand \@ifxundefined [1]{%
 \@ifx{#1\undefined}
}%
\providecommand \@ifnum [1]{%
 \ifnum #1\expandafter \@firstoftwo
 \else \expandafter \@secondoftwo
 \fi
}%
\providecommand \@ifx [1]{%
 \ifx #1\expandafter \@firstoftwo
 \else \expandafter \@secondoftwo
 \fi
}%
\providecommand \natexlab [1]{#1}%
\providecommand \enquote  [1]{``#1''}%
\providecommand \bibnamefont  [1]{#1}%
\providecommand \bibfnamefont [1]{#1}%
\providecommand \citenamefont [1]{#1}%
\providecommand \href@noop [0]{\@secondoftwo}%
\providecommand \href [0]{\begingroup \@sanitize@url \@href}%
\providecommand \@href[1]{\@@startlink{#1}\@@href}%
\providecommand \@@href[1]{\endgroup#1\@@endlink}%
\providecommand \@sanitize@url [0]{\catcode `\\12\catcode `\$12\catcode
  `\&12\catcode `\#12\catcode `\^12\catcode `\_12\catcode `\%12\relax}%
\providecommand \@@startlink[1]{}%
\providecommand \@@endlink[0]{}%
\providecommand \url  [0]{\begingroup\@sanitize@url \@url }%
\providecommand \@url [1]{\endgroup\@href {#1}{\urlprefix }}%
\providecommand \urlprefix  [0]{URL }%
\providecommand \Eprint [0]{\href }%
\providecommand \doibase [0]{https://doi.org/}%
\providecommand \selectlanguage [0]{\@gobble}%
\providecommand \bibinfo  [0]{\@secondoftwo}%
\providecommand \bibfield  [0]{\@secondoftwo}%
\providecommand \translation [1]{[#1]}%
\providecommand \BibitemOpen [0]{}%
\providecommand \bibitemStop [0]{}%
\providecommand \bibitemNoStop [0]{.\EOS\space}%
\providecommand \EOS [0]{\spacefactor3000\relax}%
\providecommand \BibitemShut  [1]{\csname bibitem#1\endcsname}%
\let\auto@bib@innerbib\@empty
\bibitem [{\citenamefont {Papadimitriou}\ and\ \citenamefont
  {Steiglitz}(1998)}]{Papadimitriou1998combinatorial}%
  \BibitemOpen
  \bibfield  {author} {\bibinfo {author} {\bibfnamefont {C.~H.}\ \bibnamefont
  {Papadimitriou}}\ and\ \bibinfo {author} {\bibfnamefont {K.}~\bibnamefont
  {Steiglitz}},\ }\href@noop {} {\emph {\bibinfo {title} {Combinatorial
  optimization: algorithms and complexity}}}\ (\bibinfo  {publisher} {Courier
  Corporation},\ \bibinfo {year} {1998})\BibitemShut {NoStop}%
\bibitem [{\citenamefont {Lucas}(2014)}]{Lucas2014ising}%
  \BibitemOpen
  \bibfield  {author} {\bibinfo {author} {\bibfnamefont {A.}~\bibnamefont
  {Lucas}},\ }\bibfield  {title} {\bibinfo {title} {Ising formulations of many
  {NP} problems},\ }\href@noop {} {\bibfield  {journal} {\bibinfo  {journal}
  {Front. Physics}\ }\textbf {\bibinfo {volume} {2}},\ \bibinfo {pages} {5}
  (\bibinfo {year} {2014})}\BibitemShut {NoStop}%
\bibitem [{\citenamefont {Sch{\"o}ning}(1999)}]{Schoening1999probabilistic}%
  \BibitemOpen
  \bibfield  {author} {\bibinfo {author} {\bibfnamefont {U.}~\bibnamefont
  {Sch{\"o}ning}},\ }\bibfield  {title} {\bibinfo {title} {A probabilistic
  algorithm for {k-SAT} and constraint satisfaction problems},\ }in\ \href@noop
  {} {\emph {\bibinfo {booktitle} {40th Annual Symposium on Foundations of
  Computer Science (Cat. No. 99CB37039)}}}\ (\bibinfo {organization} {IEEE},\
  \bibinfo {year} {1999})\ pp.\ \bibinfo {pages} {410--414}\BibitemShut
  {NoStop}%
\bibitem [{\citenamefont {Bertsimas}\ and\ \citenamefont
  {Tsitsiklis}(1993)}]{Bertsimas1993simulated}%
  \BibitemOpen
  \bibfield  {author} {\bibinfo {author} {\bibfnamefont {D.}~\bibnamefont
  {Bertsimas}}\ and\ \bibinfo {author} {\bibfnamefont {J.}~\bibnamefont
  {Tsitsiklis}},\ }\bibfield  {title} {\bibinfo {title} {{Simulated
  Annealing}},\ }\href {https://doi.org/10.1214/ss/1177011077} {\bibfield
  {journal} {\bibinfo  {journal} {Statistical Science}\ }\textbf {\bibinfo
  {volume} {8}},\ \bibinfo {pages} {10 } (\bibinfo {year} {1993})}\BibitemShut
  {NoStop}%
\bibitem [{\citenamefont {Chen}\ and\ \citenamefont
  {Aihara}(1995)}]{Chen1995chaotic}%
  \BibitemOpen
  \bibfield  {author} {\bibinfo {author} {\bibfnamefont {L.}~\bibnamefont
  {Chen}}\ and\ \bibinfo {author} {\bibfnamefont {K.}~\bibnamefont {Aihara}},\
  }\bibfield  {title} {\bibinfo {title} {Chaotic simulated annealing by a
  neural network model with transient chaos},\ }\href
  {https://doi.org/https://doi.org/10.1016/0893-6080(95)00033-V} {\bibfield
  {journal} {\bibinfo  {journal} {Neural Networks}\ }\textbf {\bibinfo {volume}
  {8}},\ \bibinfo {pages} {915} (\bibinfo {year} {1995})}\BibitemShut {NoStop}%
\bibitem [{\citenamefont {Ercsey-Ravasz}\ and\ \citenamefont
  {Toroczkai}(2011)}]{Ercsey2011optimization}%
  \BibitemOpen
  \bibfield  {author} {\bibinfo {author} {\bibfnamefont {M.}~\bibnamefont
  {Ercsey-Ravasz}}\ and\ \bibinfo {author} {\bibfnamefont {Z.}~\bibnamefont
  {Toroczkai}},\ }\bibfield  {title} {\bibinfo {title} {Optimization hardness
  as transient chaos in an analog approach to constraint satisfaction},\
  }\href@noop {} {\bibfield  {journal} {\bibinfo  {journal} {Nat. Phys.}\
  }\textbf {\bibinfo {volume} {7}},\ \bibinfo {pages} {966} (\bibinfo {year}
  {2011})}\BibitemShut {NoStop}%
\bibitem [{\citenamefont {Ercsey-Ravasz}\ and\ \citenamefont
  {Toroczkai}(2012)}]{Ercsey2012chaos}%
  \BibitemOpen
  \bibfield  {author} {\bibinfo {author} {\bibfnamefont {M.}~\bibnamefont
  {Ercsey-Ravasz}}\ and\ \bibinfo {author} {\bibfnamefont {Z.}~\bibnamefont
  {Toroczkai}},\ }\bibfield  {title} {\bibinfo {title} {The chaos within
  {S}udoku},\ }\href@noop {} {\bibfield  {journal} {\bibinfo  {journal} {Sci.
  Rep.}\ }\textbf {\bibinfo {volume} {2}},\ \bibinfo {pages} {725} (\bibinfo
  {year} {2012})}\BibitemShut {NoStop}%
\bibitem [{\citenamefont {Wang}\ \emph {et~al.}(2013)\citenamefont {Wang},
  \citenamefont {Marandi}, \citenamefont {Wen}, \citenamefont {Byer},\ and\
  \citenamefont {Yamamoto}}]{Wang2013coherent}%
  \BibitemOpen
  \bibfield  {author} {\bibinfo {author} {\bibfnamefont {Z.}~\bibnamefont
  {Wang}}, \bibinfo {author} {\bibfnamefont {A.}~\bibnamefont {Marandi}},
  \bibinfo {author} {\bibfnamefont {K.}~\bibnamefont {Wen}}, \bibinfo {author}
  {\bibfnamefont {R.~L.}\ \bibnamefont {Byer}},\ and\ \bibinfo {author}
  {\bibfnamefont {Y.}~\bibnamefont {Yamamoto}},\ }\bibfield  {title} {\bibinfo
  {title} {Coherent {I}sing machine based on degenerate optical parametric
  oscillators},\ }\href {https://doi.org/10.1103/PhysRevA.88.063853} {\bibfield
   {journal} {\bibinfo  {journal} {Phys. Rev. A}\ }\textbf {\bibinfo {volume}
  {88}},\ \bibinfo {pages} {063853} (\bibinfo {year} {2013})}\BibitemShut
  {NoStop}%
\bibitem [{\citenamefont {Yamamoto}\ \emph {et~al.}(2017)\citenamefont
  {Yamamoto}, \citenamefont {Aihara}, \citenamefont {Leleu}, \citenamefont
  {Kawarabayashi}, \citenamefont {Kako}, \citenamefont {Fejer}, \citenamefont
  {Inoue},\ and\ \citenamefont {Takesue}}]{Yamamoto2017coherent}%
  \BibitemOpen
  \bibfield  {author} {\bibinfo {author} {\bibfnamefont {Y.}~\bibnamefont
  {Yamamoto}}, \bibinfo {author} {\bibfnamefont {K.}~\bibnamefont {Aihara}},
  \bibinfo {author} {\bibfnamefont {T.}~\bibnamefont {Leleu}}, \bibinfo
  {author} {\bibfnamefont {K.-I.}\ \bibnamefont {Kawarabayashi}}, \bibinfo
  {author} {\bibfnamefont {S.}~\bibnamefont {Kako}}, \bibinfo {author}
  {\bibfnamefont {M.}~\bibnamefont {Fejer}}, \bibinfo {author} {\bibfnamefont
  {K.}~\bibnamefont {Inoue}},\ and\ \bibinfo {author} {\bibfnamefont
  {H.}~\bibnamefont {Takesue}},\ }\bibfield  {title} {\bibinfo {title}
  {Coherent {I}sing machines—optical neural networks operating at the quantum
  limit},\ }\href@noop {} {\bibfield  {journal} {\bibinfo  {journal} {npj
  Quantum Information}\ }\textbf {\bibinfo {volume} {3}},\ \bibinfo {pages}
  {49} (\bibinfo {year} {2017})}\BibitemShut {NoStop}%
\bibitem [{\citenamefont {Berloff}\ \emph {et~al.}(2017)\citenamefont
  {Berloff}, \citenamefont {Silva}, \citenamefont {Kalinin}, \citenamefont
  {Askitopoulos}, \citenamefont {T{\"o}pfer}, \citenamefont {Cilibrizzi},
  \citenamefont {Langbein},\ and\ \citenamefont
  {Lagoudakis}}]{Berloff2017realizing}%
  \BibitemOpen
  \bibfield  {author} {\bibinfo {author} {\bibfnamefont {N.~G.}\ \bibnamefont
  {Berloff}}, \bibinfo {author} {\bibfnamefont {M.}~\bibnamefont {Silva}},
  \bibinfo {author} {\bibfnamefont {K.}~\bibnamefont {Kalinin}}, \bibinfo
  {author} {\bibfnamefont {A.}~\bibnamefont {Askitopoulos}}, \bibinfo {author}
  {\bibfnamefont {J.~D.}\ \bibnamefont {T{\"o}pfer}}, \bibinfo {author}
  {\bibfnamefont {P.}~\bibnamefont {Cilibrizzi}}, \bibinfo {author}
  {\bibfnamefont {W.}~\bibnamefont {Langbein}},\ and\ \bibinfo {author}
  {\bibfnamefont {P.~G.}\ \bibnamefont {Lagoudakis}},\ }\bibfield  {title}
  {\bibinfo {title} {Realizing the classical {XY} {H}amiltonian in polariton
  simulators},\ }\href@noop {} {\bibfield  {journal} {\bibinfo  {journal} {Nat.
  Mat.}\ }\textbf {\bibinfo {volume} {16}},\ \bibinfo {pages} {1120} (\bibinfo
  {year} {2017})}\BibitemShut {NoStop}%
\bibitem [{\citenamefont {King}\ \emph {et~al.}()\citenamefont {King},
  \citenamefont {Bernoudy}, \citenamefont {King}, \citenamefont {Berkley},\
  and\ \citenamefont {Lanting}}]{King2018emulating}%
  \BibitemOpen
  \bibfield  {author} {\bibinfo {author} {\bibfnamefont {A.~D.}\ \bibnamefont
  {King}}, \bibinfo {author} {\bibfnamefont {W.}~\bibnamefont {Bernoudy}},
  \bibinfo {author} {\bibfnamefont {J.}~\bibnamefont {King}}, \bibinfo {author}
  {\bibfnamefont {A.~J.}\ \bibnamefont {Berkley}},\ and\ \bibinfo {author}
  {\bibfnamefont {T.}~\bibnamefont {Lanting}},\ }\bibfield  {title} {\bibinfo
  {title} {Emulating the coherent {I}sing machine with a mean-field
  algorithm},\ }\href@noop {} {\bibinfo  {journal} {arXiv:1806.08422}\
  }\BibitemShut {NoStop}%
\bibitem [{\citenamefont {Leleu}\ \emph {et~al.}(2019)\citenamefont {Leleu},
  \citenamefont {Yamamoto}, \citenamefont {McMahon},\ and\ \citenamefont
  {Aihara}}]{Leleu2019destabilization}%
  \BibitemOpen
\bibfield  {journal} {  }\bibfield  {author} {\bibinfo {author} {\bibfnamefont
  {T.}~\bibnamefont {Leleu}}, \bibinfo {author} {\bibfnamefont
  {Y.}~\bibnamefont {Yamamoto}}, \bibinfo {author} {\bibfnamefont {P.~L.}\
  \bibnamefont {McMahon}},\ and\ \bibinfo {author} {\bibfnamefont
  {K.}~\bibnamefont {Aihara}},\ }\bibfield  {title} {\bibinfo {title}
  {Destabilization of local minima in analog spin systems by correction of
  amplitude heterogeneity},\ }\href
  {https://doi.org/10.1103/PhysRevLett.122.040607} {\bibfield  {journal}
  {\bibinfo  {journal} {Phys. Rev. Lett.}\ }\textbf {\bibinfo {volume} {122}},\
  \bibinfo {pages} {040607} (\bibinfo {year} {2019})}\BibitemShut {NoStop}%
\bibitem [{\citenamefont {Goto}\ \emph {et~al.}(2019)\citenamefont {Goto},
  \citenamefont {Tatsumura},\ and\ \citenamefont
  {Dixon}}]{Goto2019combinatorial}%
  \BibitemOpen
  \bibfield  {author} {\bibinfo {author} {\bibfnamefont {H.}~\bibnamefont
  {Goto}}, \bibinfo {author} {\bibfnamefont {K.}~\bibnamefont {Tatsumura}},\
  and\ \bibinfo {author} {\bibfnamefont {A.~R.}\ \bibnamefont {Dixon}},\
  }\bibfield  {title} {\bibinfo {title} {Combinatorial optimization by
  simulating adiabatic bifurcations in nonlinear {H}amiltonian systems},\
  }\href@noop {} {\bibfield  {journal} {\bibinfo  {journal} {Sci. Adv.}\
  }\textbf {\bibinfo {volume} {5}},\ \bibinfo {pages} {eaav2372} (\bibinfo
  {year} {2019})}\BibitemShut {NoStop}%
\bibitem [{\citenamefont {Vadlamani}\ \emph {et~al.}(2020)\citenamefont
  {Vadlamani}, \citenamefont {Xiao},\ and\ \citenamefont
  {Yablonovitch}}]{Vadlamani2020physics}%
  \BibitemOpen
  \bibfield  {author} {\bibinfo {author} {\bibfnamefont {S.~K.}\ \bibnamefont
  {Vadlamani}}, \bibinfo {author} {\bibfnamefont {T.~P.}\ \bibnamefont
  {Xiao}},\ and\ \bibinfo {author} {\bibfnamefont {E.}~\bibnamefont
  {Yablonovitch}},\ }\bibfield  {title} {\bibinfo {title} {Physics successfully
  implements {L}agrange multiplier optimization},\ }\href@noop {} {\bibfield
  {journal} {\bibinfo  {journal} {Proc. Nat. Acad. Sci.}\ }\textbf {\bibinfo
  {volume} {117}},\ \bibinfo {pages} {26639} (\bibinfo {year}
  {2020})}\BibitemShut {NoStop}%
\bibitem [{\citenamefont {Goto}\ \emph {et~al.}(2021)\citenamefont {Goto},
  \citenamefont {Endo}, \citenamefont {Suzuki}, \citenamefont {Sakai},
  \citenamefont {Kanao}, \citenamefont {Hamakawa}, \citenamefont {Hidaka},
  \citenamefont {Yamasaki},\ and\ \citenamefont
  {Tatsumura}}]{Goto2021performance}%
  \BibitemOpen
  \bibfield  {author} {\bibinfo {author} {\bibfnamefont {H.}~\bibnamefont
  {Goto}}, \bibinfo {author} {\bibfnamefont {K.}~\bibnamefont {Endo}}, \bibinfo
  {author} {\bibfnamefont {M.}~\bibnamefont {Suzuki}}, \bibinfo {author}
  {\bibfnamefont {Y.}~\bibnamefont {Sakai}}, \bibinfo {author} {\bibfnamefont
  {T.}~\bibnamefont {Kanao}}, \bibinfo {author} {\bibfnamefont
  {Y.}~\bibnamefont {Hamakawa}}, \bibinfo {author} {\bibfnamefont
  {R.}~\bibnamefont {Hidaka}}, \bibinfo {author} {\bibfnamefont
  {M.}~\bibnamefont {Yamasaki}},\ and\ \bibinfo {author} {\bibfnamefont
  {K.}~\bibnamefont {Tatsumura}},\ }\bibfield  {title} {\bibinfo {title}
  {High-performance combinatorial optimization based on classical mechanics},\
  }\href@noop {} {\bibfield  {journal} {\bibinfo  {journal} {Sci. Adv.}\
  }\textbf {\bibinfo {volume} {7}},\ \bibinfo {pages} {eabe7953} (\bibinfo
  {year} {2021})}\BibitemShut {NoStop}%
\bibitem [{\citenamefont {Mohseni}\ \emph {et~al.}(2022)\citenamefont
  {Mohseni}, \citenamefont {McMahon},\ and\ \citenamefont
  {Byrnes}}]{Mohseni2022ising}%
  \BibitemOpen
  \bibfield  {author} {\bibinfo {author} {\bibfnamefont {N.}~\bibnamefont
  {Mohseni}}, \bibinfo {author} {\bibfnamefont {P.~L.}\ \bibnamefont
  {McMahon}},\ and\ \bibinfo {author} {\bibfnamefont {T.}~\bibnamefont
  {Byrnes}},\ }\bibfield  {title} {\bibinfo {title} {Ising machines as hardware
  solvers of combinatorial optimization problems},\ }\href@noop {} {\bibfield
  {journal} {\bibinfo  {journal} {Nat. Rev. Phys.}\ }\textbf {\bibinfo {volume}
  {4}},\ \bibinfo {pages} {363} (\bibinfo {year} {2022})}\BibitemShut {NoStop}%
\bibitem [{\citenamefont {Reifenstein}\ \emph {et~al.}(2023)\citenamefont
  {Reifenstein}, \citenamefont {Leleu}, \citenamefont {McKenna}, \citenamefont
  {Jankowski}, \citenamefont {Suh}, \citenamefont {Ng}, \citenamefont
  {Khoyratee}, \citenamefont {Toroczkai},\ and\ \citenamefont
  {Yamamoto}}]{Reifenstein2023coherent}%
  \BibitemOpen
  \bibfield  {author} {\bibinfo {author} {\bibfnamefont {S.}~\bibnamefont
  {Reifenstein}}, \bibinfo {author} {\bibfnamefont {T.}~\bibnamefont {Leleu}},
  \bibinfo {author} {\bibfnamefont {T.}~\bibnamefont {McKenna}}, \bibinfo
  {author} {\bibfnamefont {M.}~\bibnamefont {Jankowski}}, \bibinfo {author}
  {\bibfnamefont {M.-G.}\ \bibnamefont {Suh}}, \bibinfo {author} {\bibfnamefont
  {E.}~\bibnamefont {Ng}}, \bibinfo {author} {\bibfnamefont {F.}~\bibnamefont
  {Khoyratee}}, \bibinfo {author} {\bibfnamefont {Z.}~\bibnamefont
  {Toroczkai}},\ and\ \bibinfo {author} {\bibfnamefont {Y.}~\bibnamefont
  {Yamamoto}},\ }\bibfield  {title} {\bibinfo {title} {Coherent {SAT} solvers:
  a tutorial},\ }\href@noop {} {\bibfield  {journal} {\bibinfo  {journal} {Adv.
  Opt. Phot.}\ }\textbf {\bibinfo {volume} {15}},\ \bibinfo {pages} {385}
  (\bibinfo {year} {2023})}\BibitemShut {NoStop}%
\bibitem [{\citenamefont {Syed}\ and\ \citenamefont
  {Berloff}(2023)}]{Syed2023physics}%
  \BibitemOpen
  \bibfield  {author} {\bibinfo {author} {\bibfnamefont {M.}~\bibnamefont
  {Syed}}\ and\ \bibinfo {author} {\bibfnamefont {N.~G.}\ \bibnamefont
  {Berloff}},\ }\bibfield  {title} {\bibinfo {title} {Physics-enhanced
  bifurcation optimisers: all you need is a canonical complex network},\
  }\href@noop {} {\bibfield  {journal} {\bibinfo  {journal} {IEEE Journal of
  Selected Topics in Quantum Electronics}\ }\textbf {\bibinfo {volume} {29}},\
  \bibinfo {pages} {1} (\bibinfo {year} {2023})}\BibitemShut {NoStop}%
\bibitem [{\citenamefont {Sch{\"o}nhage}(1979)}]{Schoenhage1979power}%
  \BibitemOpen
  \bibfield  {author} {\bibinfo {author} {\bibfnamefont {A.}~\bibnamefont
  {Sch{\"o}nhage}},\ }\bibfield  {title} {\bibinfo {title} {On the power of
  random access machines},\ }in\ \href@noop {} {\emph {\bibinfo {booktitle}
  {International Colloquium on Automata, Languages, and Programming}}}\
  (\bibinfo {organization} {Springer},\ \bibinfo {year} {1979})\ pp.\ \bibinfo
  {pages} {520--529}\BibitemShut {NoStop}%
\bibitem [{\citenamefont {Leleu}\ \emph {et~al.}(2021)\citenamefont {Leleu},
  \citenamefont {Khoyratee}, \citenamefont {Levi}, \citenamefont {Hamerly},
  \citenamefont {Kohno},\ and\ \citenamefont {Aihara}}]{Leleu2021scaling}%
  \BibitemOpen
  \bibfield  {author} {\bibinfo {author} {\bibfnamefont {T.}~\bibnamefont
  {Leleu}}, \bibinfo {author} {\bibfnamefont {F.}~\bibnamefont {Khoyratee}},
  \bibinfo {author} {\bibfnamefont {T.}~\bibnamefont {Levi}}, \bibinfo {author}
  {\bibfnamefont {R.}~\bibnamefont {Hamerly}}, \bibinfo {author} {\bibfnamefont
  {T.}~\bibnamefont {Kohno}},\ and\ \bibinfo {author} {\bibfnamefont
  {K.}~\bibnamefont {Aihara}},\ }\bibfield  {title} {\bibinfo {title} {Scaling
  advantage of chaotic amplitude control for high-performance combinatorial
  optimization},\ }\href@noop {} {\bibfield  {journal} {\bibinfo  {journal}
  {Comm. Phys.}\ }\textbf {\bibinfo {volume} {4}},\ \bibinfo {pages} {266}
  (\bibinfo {year} {2021})}\BibitemShut {NoStop}%
\bibitem [{\citenamefont {Albash}\ \emph {et~al.}(2019)\citenamefont {Albash},
  \citenamefont {Martin-Mayor},\ and\ \citenamefont {Hen}}]{Albash2019analog}%
  \BibitemOpen
  \bibfield  {author} {\bibinfo {author} {\bibfnamefont {T.}~\bibnamefont
  {Albash}}, \bibinfo {author} {\bibfnamefont {V.}~\bibnamefont
  {Martin-Mayor}},\ and\ \bibinfo {author} {\bibfnamefont {I.}~\bibnamefont
  {Hen}},\ }\bibfield  {title} {\bibinfo {title} {Analog errors in {I}sing
  machines},\ }\href@noop {} {\bibfield  {journal} {\bibinfo  {journal}
  {Quantum Science and Technology}\ }\textbf {\bibinfo {volume} {4}},\ \bibinfo
  {pages} {02LT03} (\bibinfo {year} {2019})}\BibitemShut {NoStop}%
\bibitem [{\citenamefont {Sumi}\ \emph {et~al.}(2014)\citenamefont {Sumi},
  \citenamefont {Moln{\'a}r},\ and\ \citenamefont
  {Ercsey-Ravasz}}]{Sumi2014robust}%
  \BibitemOpen
  \bibfield  {author} {\bibinfo {author} {\bibfnamefont {R.}~\bibnamefont
  {Sumi}}, \bibinfo {author} {\bibfnamefont {B.}~\bibnamefont {Moln{\'a}r}},\
  and\ \bibinfo {author} {\bibfnamefont {M.}~\bibnamefont {Ercsey-Ravasz}},\
  }\bibfield  {title} {\bibinfo {title} {Robust optimization with transiently
  chaotic dynamical systems},\ }\href@noop {} {\bibfield  {journal} {\bibinfo
  {journal} {Europhys. Lett.}\ }\textbf {\bibinfo {volume} {106}},\ \bibinfo
  {pages} {40002} (\bibinfo {year} {2014})}\BibitemShut {NoStop}%
\bibitem [{\citenamefont {Nishimori}(2001)}]{Nishimori2001statistical}%
  \BibitemOpen
  \bibfield  {author} {\bibinfo {author} {\bibfnamefont {H.}~\bibnamefont
  {Nishimori}},\ }\href@noop {} {\emph {\bibinfo {title} {Statistical physics
  of spin glasses and information processing: an introduction}}}\ (\bibinfo
  {publisher} {Oxford University Press},\ \bibinfo {year} {2001})\BibitemShut
  {NoStop}%
\bibitem [{\citenamefont {Marandi}\ \emph {et~al.}(2014)\citenamefont
  {Marandi}, \citenamefont {Wang}, \citenamefont {Takata}, \citenamefont
  {Byer},\ and\ \citenamefont {Yamamoto}}]{Marandi2014network}%
  \BibitemOpen
  \bibfield  {author} {\bibinfo {author} {\bibfnamefont {A.}~\bibnamefont
  {Marandi}}, \bibinfo {author} {\bibfnamefont {Z.}~\bibnamefont {Wang}},
  \bibinfo {author} {\bibfnamefont {K.}~\bibnamefont {Takata}}, \bibinfo
  {author} {\bibfnamefont {R.~L.}\ \bibnamefont {Byer}},\ and\ \bibinfo
  {author} {\bibfnamefont {Y.}~\bibnamefont {Yamamoto}},\ }\bibfield  {title}
  {\bibinfo {title} {Network of time-multiplexed optical parametric oscillators
  as a coherent {I}sing machine},\ }\href@noop {} {\bibfield  {journal}
  {\bibinfo  {journal} {Nat. Phot.}\ }\textbf {\bibinfo {volume} {8}},\
  \bibinfo {pages} {937} (\bibinfo {year} {2014})}\BibitemShut {NoStop}%
\bibitem [{\citenamefont {McMahon}\ \emph {et~al.}(2016)\citenamefont
  {McMahon}, \citenamefont {Marandi}, \citenamefont {Haribara}, \citenamefont
  {Hamerly}, \citenamefont {Langrock}, \citenamefont {Tamate}, \citenamefont
  {Inagaki}, \citenamefont {Takesue}, \citenamefont {Utsunomiya}, \citenamefont
  {Aihara}, \citenamefont {Byer}, \citenamefont {Fejer}, \citenamefont
  {Mabuchi},\ and\ \citenamefont {Yamamoto}}]{Mcmahon2016fully}%
  \BibitemOpen
  \bibfield  {author} {\bibinfo {author} {\bibfnamefont {P.~L.}\ \bibnamefont
  {McMahon}}, \bibinfo {author} {\bibfnamefont {A.}~\bibnamefont {Marandi}},
  \bibinfo {author} {\bibfnamefont {Y.}~\bibnamefont {Haribara}}, \bibinfo
  {author} {\bibfnamefont {R.}~\bibnamefont {Hamerly}}, \bibinfo {author}
  {\bibfnamefont {C.}~\bibnamefont {Langrock}}, \bibinfo {author}
  {\bibfnamefont {S.}~\bibnamefont {Tamate}}, \bibinfo {author} {\bibfnamefont
  {T.}~\bibnamefont {Inagaki}}, \bibinfo {author} {\bibfnamefont
  {H.}~\bibnamefont {Takesue}}, \bibinfo {author} {\bibfnamefont
  {S.}~\bibnamefont {Utsunomiya}}, \bibinfo {author} {\bibfnamefont
  {K.}~\bibnamefont {Aihara}}, \bibinfo {author} {\bibfnamefont {R.~L.}\
  \bibnamefont {Byer}}, \bibinfo {author} {\bibfnamefont {M.~M.}\ \bibnamefont
  {Fejer}}, \bibinfo {author} {\bibfnamefont {H.}~\bibnamefont {Mabuchi}},\
  and\ \bibinfo {author} {\bibfnamefont {Y.}~\bibnamefont {Yamamoto}},\
  }\bibfield  {title} {\bibinfo {title} {A fully programmable 100-spin coherent
  {I}sing machine with all-to-all connections},\ }\href@noop {} {\bibfield
  {journal} {\bibinfo  {journal} {Science}\ }\textbf {\bibinfo {volume}
  {354}},\ \bibinfo {pages} {614} (\bibinfo {year} {2016})}\BibitemShut
  {NoStop}%
\bibitem [{\citenamefont {Inagaki}\ \emph {et~al.}(2016)\citenamefont
  {Inagaki}, \citenamefont {Haribara}, \citenamefont {Igarashi}, \citenamefont
  {Sonobe}, \citenamefont {Tamate}, \citenamefont {Honjo}, \citenamefont
  {Marandi}, \citenamefont {McMahon}, \citenamefont {Umeki}, \citenamefont
  {Enbutsu}, \citenamefont {Tadanaga}, \citenamefont {Takenouchi},
  \citenamefont {Aihara}, \citenamefont {Kawarabayashi}, \citenamefont {Inoue},
  \citenamefont {Utsunomiya},\ and\ \citenamefont
  {Takesue}}]{Inagaki2016coherent}%
  \BibitemOpen
  \bibfield  {author} {\bibinfo {author} {\bibfnamefont {T.}~\bibnamefont
  {Inagaki}}, \bibinfo {author} {\bibfnamefont {Y.}~\bibnamefont {Haribara}},
  \bibinfo {author} {\bibfnamefont {K.}~\bibnamefont {Igarashi}}, \bibinfo
  {author} {\bibfnamefont {T.}~\bibnamefont {Sonobe}}, \bibinfo {author}
  {\bibfnamefont {S.}~\bibnamefont {Tamate}}, \bibinfo {author} {\bibfnamefont
  {T.}~\bibnamefont {Honjo}}, \bibinfo {author} {\bibfnamefont
  {A.}~\bibnamefont {Marandi}}, \bibinfo {author} {\bibfnamefont {P.~L.}\
  \bibnamefont {McMahon}}, \bibinfo {author} {\bibfnamefont {T.}~\bibnamefont
  {Umeki}}, \bibinfo {author} {\bibfnamefont {K.}~\bibnamefont {Enbutsu}},
  \bibinfo {author} {\bibfnamefont {O.}~\bibnamefont {Tadanaga}}, \bibinfo
  {author} {\bibfnamefont {H.}~\bibnamefont {Takenouchi}}, \bibinfo {author}
  {\bibfnamefont {K.}~\bibnamefont {Aihara}}, \bibinfo {author} {\bibfnamefont
  {K.}~\bibnamefont {Kawarabayashi}}, \bibinfo {author} {\bibfnamefont
  {K.}~\bibnamefont {Inoue}}, \bibinfo {author} {\bibfnamefont
  {S.}~\bibnamefont {Utsunomiya}},\ and\ \bibinfo {author} {\bibfnamefont
  {H.}~\bibnamefont {Takesue}},\ }\bibfield  {title} {\bibinfo {title} {A
  coherent {I}sing machine for 2000-node optimization problems},\ }\href@noop
  {} {\bibfield  {journal} {\bibinfo  {journal} {Science}\ }\textbf {\bibinfo
  {volume} {354}},\ \bibinfo {pages} {603} (\bibinfo {year}
  {2016})}\BibitemShut {NoStop}%
\bibitem [{\citenamefont {Yamamoto}\ \emph {et~al.}(2020)\citenamefont
  {Yamamoto}, \citenamefont {Leleu}, \citenamefont {Ganguli},\ and\
  \citenamefont {Mabuchi}}]{Yamamoto2020coherent}%
  \BibitemOpen
  \bibfield  {author} {\bibinfo {author} {\bibfnamefont {Y.}~\bibnamefont
  {Yamamoto}}, \bibinfo {author} {\bibfnamefont {T.}~\bibnamefont {Leleu}},
  \bibinfo {author} {\bibfnamefont {S.}~\bibnamefont {Ganguli}},\ and\ \bibinfo
  {author} {\bibfnamefont {H.}~\bibnamefont {Mabuchi}},\ }\bibfield  {title}
  {\bibinfo {title} {Coherent {I}sing machines—quantum optics and neural
  network perspectives},\ }\href@noop {} {\bibfield  {journal} {\bibinfo
  {journal} {Appl. Phys. Lett.}\ }\textbf {\bibinfo {volume} {117}} (\bibinfo
  {year} {2020})}\BibitemShut {NoStop}%
\bibitem [{\citenamefont {Honjo}\ \emph {et~al.}(2021)\citenamefont {Honjo},
  \citenamefont {Sonobe}, \citenamefont {Inaba}, \citenamefont {Inagaki},
  \citenamefont {Ikuta}, \citenamefont {Yamada}, \citenamefont {Kazama},
  \citenamefont {Enbutsu}, \citenamefont {Umeki}, \citenamefont {Kasahara},
  \citenamefont {Kawarabayashi},\ and\ \citenamefont
  {Takesue}}]{Honjo2021spin}%
  \BibitemOpen
  \bibfield  {author} {\bibinfo {author} {\bibfnamefont {T.}~\bibnamefont
  {Honjo}}, \bibinfo {author} {\bibfnamefont {T.}~\bibnamefont {Sonobe}},
  \bibinfo {author} {\bibfnamefont {K.}~\bibnamefont {Inaba}}, \bibinfo
  {author} {\bibfnamefont {T.}~\bibnamefont {Inagaki}}, \bibinfo {author}
  {\bibfnamefont {T.}~\bibnamefont {Ikuta}}, \bibinfo {author} {\bibfnamefont
  {Y.}~\bibnamefont {Yamada}}, \bibinfo {author} {\bibfnamefont
  {T.}~\bibnamefont {Kazama}}, \bibinfo {author} {\bibfnamefont
  {K.}~\bibnamefont {Enbutsu}}, \bibinfo {author} {\bibfnamefont
  {T.}~\bibnamefont {Umeki}}, \bibinfo {author} {\bibfnamefont
  {R.}~\bibnamefont {Kasahara}}, \bibinfo {author} {\bibfnamefont
  {K.}~\bibnamefont {Kawarabayashi}},\ and\ \bibinfo {author} {\bibfnamefont
  {H.}~\bibnamefont {Takesue}},\ }\bibfield  {title} {\bibinfo {title}
  {100,000-spin coherent {I}sing machine},\ }\href@noop {} {\bibfield
  {journal} {\bibinfo  {journal} {Sci. Adv.}\ }\textbf {\bibinfo {volume}
  {7}},\ \bibinfo {pages} {eabh0952} (\bibinfo {year} {2021})}\BibitemShut
  {NoStop}%
\bibitem [{\citenamefont {Zhou}\ \emph {et~al.}(2023)\citenamefont {Zhou},
  \citenamefont {Gneiting}, \citenamefont {You},\ and\ \citenamefont
  {Nori}}]{Zhou2023coherent}%
  \BibitemOpen
  \bibfield  {author} {\bibinfo {author} {\bibfnamefont {Z.-Y.}\ \bibnamefont
  {Zhou}}, \bibinfo {author} {\bibfnamefont {C.}~\bibnamefont {Gneiting}},
  \bibinfo {author} {\bibfnamefont {J.~Q.}\ \bibnamefont {You}},\ and\ \bibinfo
  {author} {\bibfnamefont {F.}~\bibnamefont {Nori}},\ }\bibfield  {title}
  {\bibinfo {title} {Coherent-cluster-state generation in networks of
  degenerate optical parametric oscillators},\ }\href
  {https://doi.org/10.1103/PhysRevA.108.023704} {\bibfield  {journal} {\bibinfo
   {journal} {Phys. Rev. A}\ }\textbf {\bibinfo {volume} {108}},\ \bibinfo
  {pages} {023704} (\bibinfo {year} {2023})}\BibitemShut {NoStop}%
\bibitem [{\citenamefont {Zhou}\ \emph {et~al.}(2025)\citenamefont {Zhou},
  \citenamefont {Gneiting}, \citenamefont {You},\ and\ \citenamefont
  {Nori}}]{Zhou2024frustration}%
  \BibitemOpen
  \bibfield  {author} {\bibinfo {author} {\bibfnamefont {Z.-Y.}\ \bibnamefont
  {Zhou}}, \bibinfo {author} {\bibfnamefont {C.}~\bibnamefont {Gneiting}},
  \bibinfo {author} {\bibfnamefont {J.~Q.}\ \bibnamefont {You}},\ and\ \bibinfo
  {author} {\bibfnamefont {F.}~\bibnamefont {Nori}},\ }\bibfield  {title}
  {\bibinfo {title} {Frustration elimination and excited state search in
  coherent {I}sing machines},\ }\href
  {https://doi.org/10.1103/PhysRevLett.134.090401} {\bibfield  {journal}
  {\bibinfo  {journal} {Phys. Rev. Lett.}\ }\textbf {\bibinfo {volume} {134}},\
  \bibinfo {pages} {090401} (\bibinfo {year} {2025})}\BibitemShut {NoStop}%
\bibitem [{\citenamefont {Kako}\ \emph {et~al.}(2020)\citenamefont {Kako},
  \citenamefont {Leleu}, \citenamefont {Inui}, \citenamefont {Khoyratee},
  \citenamefont {Reifenstein},\ and\ \citenamefont
  {Yamamoto}}]{Kako2020coherent}%
  \BibitemOpen
  \bibfield  {author} {\bibinfo {author} {\bibfnamefont {S.}~\bibnamefont
  {Kako}}, \bibinfo {author} {\bibfnamefont {T.}~\bibnamefont {Leleu}},
  \bibinfo {author} {\bibfnamefont {Y.}~\bibnamefont {Inui}}, \bibinfo {author}
  {\bibfnamefont {F.}~\bibnamefont {Khoyratee}}, \bibinfo {author}
  {\bibfnamefont {S.}~\bibnamefont {Reifenstein}},\ and\ \bibinfo {author}
  {\bibfnamefont {Y.}~\bibnamefont {Yamamoto}},\ }\bibfield  {title} {\bibinfo
  {title} {Coherent {I}sing machines with error correction feedback},\
  }\href@noop {} {\bibfield  {journal} {\bibinfo  {journal} {Advanced Quantum
  Technologies}\ }\textbf {\bibinfo {volume} {3}},\ \bibinfo {pages} {2000045}
  (\bibinfo {year} {2020})}\BibitemShut {NoStop}%
\bibitem [{\citenamefont {Yamamura}\ \emph {et~al.}(2024)\citenamefont
  {Yamamura}, \citenamefont {Mabuchi},\ and\ \citenamefont
  {Ganguli}}]{Yamamura2024geometric}%
  \BibitemOpen
  \bibfield  {author} {\bibinfo {author} {\bibfnamefont {A.}~\bibnamefont
  {Yamamura}}, \bibinfo {author} {\bibfnamefont {H.}~\bibnamefont {Mabuchi}},\
  and\ \bibinfo {author} {\bibfnamefont {S.}~\bibnamefont {Ganguli}},\
  }\bibfield  {title} {\bibinfo {title} {Geometric landscape annealing as an
  optimization principle underlying the coherent {I}sing machine},\ }\href
  {https://doi.org/10.1103/PhysRevX.14.031054} {\bibfield  {journal} {\bibinfo
  {journal} {Phys. Rev. X}\ }\textbf {\bibinfo {volume} {14}},\ \bibinfo
  {pages} {031054} (\bibinfo {year} {2024})}\BibitemShut {NoStop}%
\bibitem [{\citenamefont {Leleu}\ \emph {et~al.}(2017)\citenamefont {Leleu},
  \citenamefont {Yamamoto}, \citenamefont {Utsunomiya},\ and\ \citenamefont
  {Aihara}}]{Leleu2017combinatorial}%
  \BibitemOpen
  \bibfield  {author} {\bibinfo {author} {\bibfnamefont {T.}~\bibnamefont
  {Leleu}}, \bibinfo {author} {\bibfnamefont {Y.}~\bibnamefont {Yamamoto}},
  \bibinfo {author} {\bibfnamefont {S.}~\bibnamefont {Utsunomiya}},\ and\
  \bibinfo {author} {\bibfnamefont {K.}~\bibnamefont {Aihara}},\ }\bibfield
  {title} {\bibinfo {title} {Combinatorial optimization using dynamical phase
  transitions in driven-dissipative systems},\ }\href
  {https://doi.org/10.1103/PhysRevE.95.022118} {\bibfield  {journal} {\bibinfo
  {journal} {Phys. Rev. E}\ }\textbf {\bibinfo {volume} {95}},\ \bibinfo
  {pages} {022118} (\bibinfo {year} {2017})}\BibitemShut {NoStop}%
\bibitem [{\citenamefont {Moln{\'a}r}\ and\ \citenamefont
  {Ercsey-Ravasz}(2013)}]{Molnar2013asymmetric}%
  \BibitemOpen
  \bibfield  {author} {\bibinfo {author} {\bibfnamefont {B.}~\bibnamefont
  {Moln{\'a}r}}\ and\ \bibinfo {author} {\bibfnamefont {M.}~\bibnamefont
  {Ercsey-Ravasz}},\ }\bibfield  {title} {\bibinfo {title} {Asymmetric
  continuous-time neural networks without local traps for solving constraint
  satisfaction problems},\ }\href@noop {} {\bibfield  {journal} {\bibinfo
  {journal} {PloS One}\ }\textbf {\bibinfo {volume} {8}},\ \bibinfo {pages}
  {e73400} (\bibinfo {year} {2013})}\BibitemShut {NoStop}%
\bibitem [{\citenamefont {Moln{\'a}r}\ \emph {et~al.}(2018)\citenamefont
  {Moln{\'a}r}, \citenamefont {Moln{\'a}r}, \citenamefont {Varga},
  \citenamefont {Toroczkai},\ and\ \citenamefont
  {Ercsey-Ravasz}}]{Molnar2018continuous}%
  \BibitemOpen
  \bibfield  {author} {\bibinfo {author} {\bibfnamefont {B.}~\bibnamefont
  {Moln{\'a}r}}, \bibinfo {author} {\bibfnamefont {F.}~\bibnamefont
  {Moln{\'a}r}}, \bibinfo {author} {\bibfnamefont {M.}~\bibnamefont {Varga}},
  \bibinfo {author} {\bibfnamefont {Z.}~\bibnamefont {Toroczkai}},\ and\
  \bibinfo {author} {\bibfnamefont {M.}~\bibnamefont {Ercsey-Ravasz}},\
  }\bibfield  {title} {\bibinfo {title} {A continuous-time {MaxSAT} solver with
  high analog performance},\ }\href@noop {} {\bibfield  {journal} {\bibinfo
  {journal} {Nat. Comm.}\ }\textbf {\bibinfo {volume} {9}},\ \bibinfo {pages}
  {4864} (\bibinfo {year} {2018})}\BibitemShut {NoStop}%
\bibitem [{\citenamefont {Moln{\'a}r}\ \emph {et~al.}(2020)\citenamefont
  {Moln{\'a}r}, \citenamefont {Kharel}, \citenamefont {Hu},\ and\ \citenamefont
  {Toroczkai}}]{Molnar2020accelerating}%
  \BibitemOpen
  \bibfield  {author} {\bibinfo {author} {\bibfnamefont {F.}~\bibnamefont
  {Moln{\'a}r}}, \bibinfo {author} {\bibfnamefont {S.~R.}\ \bibnamefont
  {Kharel}}, \bibinfo {author} {\bibfnamefont {X.~S.}\ \bibnamefont {Hu}},\
  and\ \bibinfo {author} {\bibfnamefont {Z.}~\bibnamefont {Toroczkai}},\
  }\bibfield  {title} {\bibinfo {title} {Accelerating a continuous-time analog
  {SAT} solver using {GPUs}},\ }\href@noop {} {\bibfield  {journal} {\bibinfo
  {journal} {Comp. Phys. Comm.}\ }\textbf {\bibinfo {volume} {256}},\ \bibinfo
  {pages} {107469} (\bibinfo {year} {2020})}\BibitemShut {NoStop}%
\end{thebibliography}%

\end{document}